\documentclass[letterpaper,twocolumn,10pt]{article}
\usepackage{usenix2019_v3}

\usepackage{subcaption}
\usepackage{multirow}
\usepackage{tabularx}
\usepackage{algorithmic}
\usepackage{algorithm} 

\usepackage{tikz}
\usepackage{amsmath}
\usepackage{amssymb}

\usepackage{filecontents}
\begin{document}
%-------------------------------------------------------------------------------

%don't want date printed
\date{}

\title{\Large \bf AdaDoS: Adaptive DoS Attack via Deep Adversarial Reinforcement Learning in SDN}

%for single author (just remove % characters)
\author{
{\rm Wei Shao}$^{1}$\thanks{Equal contribution.},
{\rm Yuhao Wang}$^{2}$\footnotemark[1],
{\rm Rongguang He}$^{3}$\footnotemark[1],
{\rm Muhammad Ejaz Ahmed}$^{1}$,
{\rm Seyit Camtepe}$^{1}$\\
$^{1}$Data61, CSIRO \quad
$^{2}$National University of Singapore \quad
$^{3}$Alibaba Group\\
\texttt{phdweishao@gmail.com, wangyuhao@u.nus.edu, rongguang.he@outlook.com,}\\
\texttt{Ejaz.Ahmed@data61.csiro.au, Seyit.Camtepe@data61.csiro.au}
}

 % end author

\maketitle

%-------------------------------------------------------------------------------
\begin{abstract}
%-------------------------------------------------------------------------------
Existing defence mechanisms have demonstrated significant effectiveness in mitigating rule-based Denial-of-Service (DoS) like attacks, leveraging predefined signatures and static heuristics to identify and block malicious traffic. However, the emergence of AI-driven techniques presents new challenges to SDN security, potentially compromising the efficacy of existing defence mechanisms. In this paper, we introduce~\textit{AdaDoS}, an adaptive attack model that disrupt network operations while evading detection by existing DoS-based detectors through adversarial reinforcement learning (RL). Specifically, AdaDoS models the problem as a competitive game between an attacker, whose goal is to obstruct network traffic without being detected, and a detector, which aims to identify malicious traffic. AdaDoS can solve this game by dynamically adjusting its attack strategy based on feedback from the SDN and the detector. Additionally, recognising that attackers typically have less information than defenders, AdaDoS formulates the DoS-like attack as a partially observed Markov decision process (POMDP), with the attacker having access only to delay information between attacker and victim nodes. We address this challenge with a novel reciprocal learning module, where the student agent, with limited observations, enhances its performance by learning from the teacher agent, who has full observational capabilities in the SDN environment. AdaDoS represents the first application of RL to develop DoS-like attack sequences, capable of adaptively evading both machine learning-based and rule-based DoS-like attack detectors.
\end{abstract}

%-------------------------------------------------------------------------------
\section{Introduction}
% The imperative for robust security in SDN environments is highlighted by the significant economic and operational risks associated with potential infrastructure breaches. For instance, a compromised SDN controller could enable attackers to reroute network traffic or completely disable networks, resulting in significant disruptions reminiscent of the severe economic and operational impacts seen in major Distributed Denial-of-Service (DDoS) attacks. According to a report by Kaspersky Labs, the average financial cost of a DDoS attack may reach up to $\$2$ million per incident, with operational downtime costs adding approximately $\$22,000$ per minute~\cite{reuters2021russia}. Moreover, the societal repercussions of security breaches within SDN infrastructures are profound, leading to diminished public trust and other intangible losses. For example, in the Philippines, data breaches have elicited strong public reactions, with $21\%$ of affected individuals ceasing interactions with the involved organisations and $18\%$ taking to social media to voice their concerns~\cite{unisys2023philippines}. Similarly, in the financial sector, data breaches have led to prolonged recovery periods and significant reputation damage, deterring customers from engaging with digital services and severely undermining trust in financial institutions. As a result, the banking and finance industries have been compelled to substantially increase their investment in security measures to restore consumer confidence and avert future breaches~\cite{securityintelligence2023banking}.

A Denial-of-Service (DoS) attack overwhelms network resources with illegitimate requests, blocking legitimate access. Experts in the field of cyber security have developed numerous techniques to identify and mitigate DoS attacks. Low-rate Denial of Service (LDoS) attacks are a more insidious variant of the conventional DoS, characterised by sending intermittent bursts of traffic at a low rate which is difficult to detect. Unlike traditional DoS attacks that flood the network with an overwhelming volume of traffic continuously, LDoS attacks exploit the vulnerability of specific network protocols or devices by intermittently sending packets at a rate that avoids triggering typical network security defences. This can lead to periodic service disruptions that are hard to trace and mitigate. Nevertheless, with the development of machine learning and deep learning techniques, DoS-like attack like LDoS with regular patterns can also be easily detected. For example, GASF-IPP~\cite{gasf-ipp} and FSS-RSR-based approach~\cite{tang2023detection} represent two of the latest popular approaches for detecting and mitigating LDoS attacks. Both methods utilise specifically designed features to analyse network traffic within SDN. They employ machine learning techniques to classify whether the traffic originates from an attacker or a normal node. Following this classification, they propose mitigation solutions, such as blocking suspicious IP addresses. While LDoS attacks are subtler and more resource-efficient, enabling prolonged disruptions without triggering standard high-traffic detection systems like traditional DoS attacks, their drawbacks—such as reliance on fixed patterns and limited adaptability—mean they can still be easily detected by specifically designed detectors.

\noindent\textbf{Limitations and Challenges.}
In summary, existing DoS-like attack approaches face the following challenges: (1) \textbf{Low Survival Ability:} 
DoS-like attacks with regular patterns and static attack rhythm are usually easily detectable. The structured and rhythmic nature of these attacks can inadvertently reveal their presence and result in low survival ability. This regularity is a fundamental flaw from an attacker's perspective, as it does not exploit the full potential of stealth and unpredictability that could make the attacks more effective. By adhering to these recognisable patterns, these attacks limit their ability to sustain impact without detection, reducing their overall efficacy in disrupting network operations. 
(2) \textbf{Limited Observability and Configuration Dependence:} The competition between DoS-like attackers and defenders can be conceptualised as an asymmetric game. Generally, defenders possess significant advantages over attackers due to their broader access to observational data. For example, SDN detector can observe the TCP and UDP traffic and also knows the topology and configuration of network. 
Conversely, critical knowledge needed by attackers is often not readily accessible in real-world scenarios. For example, typical LDoS attack exploits the TCP Re-transmission Timeout (RTO) intervals, which is decided by congestion control configurations and usually unknown to attackers. Besides, limited observation about network status makes attacker hard to launch efficient attack and evade detection.

% (3) \textbf{High Resource Consumption:}  DoS-like attacks inherently involve the consumption of resources, targeting the depletion of network bandwidth, server capacity, and other critical infrastructures. Although LDoS attacks are more efficient in terms of resource usage compared to traditional DoS attacks, they still face challenges in precisely modulate the consumption of resources. 

\noindent\textbf{AdaDoS - An Adaptive DoS-like Attacker.} To address the aforementioned challenges, we propose an adversarial reinforcement learning-based adaptive DoS-like attack framework under partial observable SDN environment, termed AdaDoS. 
The sequential nature of DoS attacks makes them inherently well-suited for reinforcement learning techniques, which can significantly enhance the adaptability of AdaDoS in complex and dynamic SDN environments. 

Specifically, we propose a novel adversarial reinforcement learning-based framework to dynamically adjust the attack patterns and rhythm to avoid SDN detector. By continually evolving the attack patterns, our model aims to remain elusive to static detectors. Even if a detector is trained with historical data from our attacks, our approach can evolve, ensuring that it remains difficult to detect and neutralise. Additionally, we propose a two-stage process consists of two deep neural networks: decider network, and shaper network. The decider network determines the optimal time to attack and shaper network shape the attack rate and duration. Compare to the single stage model, our proposed two-stage process make the attacks more adaptive to changing network conditions or defensive measure, and by having a separate module (the shaper) to control the attack period and rate, we can fine-tune the intensity and timing of attacks, potentially making it harder to detect and mitigate.

Additionally, to address the observation limitations, we introduce a novel transfer learning model that incorporates a teacher-student module. This model consists of a teacher agent with access to global SDN information and a student agent limited to partial observations (such as delay information between attacker and victim nodes) within real-world environment settings. Typically, the teacher agent is expected to outperform the student agent in identical scenarios due to its broader informational access. Our approach is designed for the teacher to mentor the student, allowing the student agent to refine its strategy to mirror the teacher’s, especially if the teacher achieves a higher reward in the same situations. Moreover, this model includes a reciprocal learning mechanism, where the teacher can also learn from the student if the student’s performance surpasses expectations, thereby enhancing the adaptability and effectiveness of both agents. Crucially, our approach leverages delay information, easily obtainable through $ping$ operations, which addresses the problem of limited observation and reduces dependency on specific, hard-to-acquire knowledge from the SDN context, such as details about retransmission mechanism and intrinsic latency.

% Lastly, to minimise resource consumption, we have meticulously designed a multi-objective reward function that addresses three distinct and conflicting goals: maximising the success rate to avoid detection, minimising the resources required for the attack, and maximising the effectiveness of traffic disruption. This reward function enables the framework to dynamically adjust strategies, reducing resource usage while ensuring the efficacy of the attack is maintained.

\noindent\textbf{Contributions.} This work aims to heighten awareness among the SDN security research community and industry vendors regarding potential vulnerabilities, encouraging a proactive approach to enhance defence mechanisms in SDN environments. Our contributions are outlined as follows:
\begin{itemize}
    \item We propose AdaDoS, a novel adaptive attack approach leveraging adversarial reinforcement learning to dynamically adjust the attack strategy in response to feedback from SDN environment in real time, enabling it to elude detection by both rule-based and machine learning-based security systems.
    \item We propose a novel two-stage model that determines the optimal timing and adjusts the intensity of attacks. This model utilises a carefully designed reward function to evade detection and improve the survival ability. 
    \item We design a teacher-student structure based transfer learning framework to enhance the performance of the attacker with partial observation and adaptive to various SDN environment without prior knowledge about network configuration and topology. To further improve the efficacy of this framework, we introduce a novel learning mechanism that facilitates reciprocal learning between the teacher and student agents.
    \item We conducted extensive experiments to validate the generality, robustness and performance of the AdaDoS compared to previous DoS-like attack. AdaDoS significantly outperforms the baselines with various settings. 
\end{itemize}

\section{Related Work}
In this section, we review recent advancements in the study of DoS-like attacks and corresponding defence mechanisms within SDN.

\subsection{DoS-like Attacks in SDN}
DDoS attacks overwhelm control and data plane resources by generating numerous fake data streams, causing flow table misses in OpenFlow switches and overloading controllers, thereby disrupting legitimate host communications.~\cite{shang2017flooddefender}. 
Rapid buffer filling due to these streams intensifies control plane  bandwidth strain and delays rule installations. Overflow of the forwarding table (packet forwarding rule repository managed by controller~\cite{ahmad2015security}) in OpenFlow switches can also cause issues. As numerous new forwarding rules pile up, table saturation ensues,  preventing the switch from processing further rules from the controller,  culminating in packet discards, error reports, and forwarding  latency/dropouts until table memory frees up~\cite{wang2015floodguard}.

Although DDoS attacks are renowned for their widespread impact and intense traffic deluge, LDoS attacks demonstrate a more covert and equally destructive approach, achieving their goals by precisely controlling the timing and intensity of the attack. Cao et al.~\cite{cao2019crosspath} revealed that exploiting shared links between control and data traffic disrupts SDN control channels, degrading app performance and causing severe network anomalies. Data plane attacks encompass flow table and TCP congestion  control based ones. Cao et al. introduced LOFT~\cite{cao2018disrupting}, calculating minimum attack rates to induce flow table overflow efficiently. Pascoal et al.~\cite{pascoal2020slow} detailed two LDoS tactics: one exhausts TCAM memory by continuously installing uncertain active rules, inhibiting legit rule setup; the other merges slow TCAM with low-rate saturation attacks, using minimal resources to deny service and potentially 
affecting existing legit flow rules more severely.

\subsection{DoS-like Attack Defences in SDN}
SDN solutions to address flow table overflow involve caching unmatched flows in OpenFlow switches' cache~\cite{Wang2015/12}, QoS-aware strategies~\cite{2019Defending}, and online routing schemes~\cite{2017STAR}
that optimise flow table usage. Zhao et al.~\cite{2018Joint} optimised group/flow 
tables using aggregation rules. To safeguard against DDoS, Hsu et al.~\cite{hsu2015design} suggested hashing for queue allocation and scheduling. 
Sattar et al.~\cite{sattar2016adaptive} proposed ABB for dynamic resource allocation under attacks. SDN frameworks like FleXam~\cite{shirali2013efficient} and FloodGuard~\cite{wang2015floodguard} provide packet-level control and protection against DoS. Hu et al.~\cite{hu2015comprehensive} proposed a comprehensive security architecture.
Other solutions employ statistical analysis to secure SDN controllers. Giotis et al.~\cite{giotis2016scalable} use edge switches to 
detect anomalies. Dong Ping et al.'s method~\cite{dongping16} classifies flows based on stats to spot low-traffic ones impacting control planes. 
Thomas et al.~\cite{thomas2017ddos} leverage third-party apps for filtering and DDoS detection. ML-based approaches train 
on non-attack flows: Meti et al.~\cite{meti2017detection} and Ezekiel et al.~\cite{ezekiel2017dynamic} use SVM and NN 
for DDoS detection, while Prakash et al.~\cite{prakash2018intelligent} apply KNN. Phan et al.~\cite{phan2016novel} combined SVM 
and SOM in SVMs-SOM for enhanced traffic classification and DDoS prevention.

Defence methods against LDoS attacks in SDN mainly fall into anomaly and feature detection. Anomaly detection, exemplified by ASNNC-OFA~\cite{li2021ldos}, involves clustering traffic characteristics and using outlier analysis to flag abnormalities compared to typical networks. Tang et al.~\cite{tang2022new} and Xiang Yang et al.~\cite{5696753} devised entropy and info-distance metrics to distinguish attack from legitimate traffic, while Wu et al.~\cite{wu2019sequence}
applied Smith-Waterman alignment to identify LDoS by comparing detection sequences' pulse attributes against background flows.
Feature detection revolves around extracting unique identifiers such as LDoS periodic TCP/UDP pulses. 
ADAR by X. Xu et al.~\cite{xie2020research} detects and mitigates LDoS attacks targeting SDN switch buffers. 
Tang et al. developed P\&F framework categorising traffic features into performance and attribute types~\cite{tang2021performance}. P\&F employs ML to assess if LDoS impacts normal traffic and uses time-frequency analysis to pinpoint attackers and victims. 
Their GASF-IPP framework analyses traffic with GASF transformations and UDP stream bursts within a sliding window to detect attacks~\cite{gasf-ipp}. The FSS-RSR-based scheme extracts traffic features, employs FSS-RSR for real-time LDoS detection, 
monitors traffic fluctuations to locate attacker IPs, and recommends rules to mitigate attacks~\cite{tang2023detection}.

\section{Background}
In this section, we introduce the latest detectors for Low-rate Denial of Service (LDoS) attacks within SDN environments, and the state-of-the-art in feature engineering for LDoS attack detection

\subsection{LDoS Attack}
LDoS attack is a specially designed network attack pattern that disrupts the availability of target services through periodic bursts of traffic. Unlike traditional sustained high traffic DoS attacks, the key features of LDoS attacks are their low-frequency, high-intensity traffic packets, and the interval time between attack activities. These three parameters - the duration of the attack, the attack rate, and the period between two attacks - collectively determine the effectiveness and concealment of LDoS attacks. A short duration of the attack and a long interval make these attacks harder to detect with standard traffic monitoring tools. Therefore, LDoS attacks can effectively consume the processing power and bandwidth of the target network, especially for critical applications that require stable connections, such as the TCP protocol. Even brief network delays can cause performance problems or service interruptions.

\subsection{Detector}
In SDN environments, effectively detecting LDoS attacks is crucial due to their subtle nature, which can evade conventional detection systems. Typically, SDN-based security frameworks monitor global data plane traffic through the control plane to identify, filter, and disrupt malicious flows based on distinct traffic patterns. However, the stealthy characteristics of LDoS attacks, which involve sending intermittent bursts of traffic to disrupt service without triggering standard thresholds, demand more sophisticated detection methods.

A promising approach for enhancing the detection of LDoS attacks involves the use of the Gramian Angular Summation Field (GASF), a method originally proposed by Wang et al.~\cite{wang2015imaging} for processing one-dimensional time series data. Utilising the principles of the Gram matrix, GASF excels in capturing the intricate temporal characteristics of TCP/UDP traffic flows, which is vital for recognising the anomalous patterns indicative of LDoS attacks. The strength of GASF lies in its capability to analyse both the elemental properties and the interrelationships within the data sequences, providing a granular and nuanced analysis that is well-suited for identifying the subtle fluctuations characteristic of DoS-like traffic.

Let $X(t)$ represents the traffic data reading at time $t$
Assuming a sliding window contains $l$ traffic data, as shown in equation~\ref{stream}.

\begin{equation}
    W(t) = \{(t-l+1), X(t-l+2), \ldots, X(t)\}.
    \label{stream}
\end{equation}

% \begin{equation}
%   S = s_1, s_2, \dots, s_n
%   \label{stream}
% \end{equation}

We subsequently utilise the MinMaxScaler~\cite{ScikitLearn} to normalise the stream sequence to a range between -1 and 1. The normalisation formula is provided in Equation~\ref{minmaxscaler}.

% \begin{equation}
%   s'_i=\frac{(s_i - S_{min})\times(max - min)+min}{S_{max} - S_{min}}
%   \label{minmaxscaler}
% \end{equation}

\begin{equation}
  X'(t) = \frac{(X(t) - X_{\text{min}}) \times (1 - (-1))}{X_{\text{max}} - X_{\text{min}}} + (-1),
  \label{minmaxscaler}
\end{equation}
where $X(t)$ represents the original data at time $t$, $X_\text{min}$ and $X_\text{max}$ are the minimum and maximum values in the sliding window $W(t)$. This structured approach ensures the data is scaled appropriately, keeping the integrity and relational properties of the original data.

% The range of $i$ spans from $1$ to $n$. $S_{min}$ and $S_{max}$ denote the minimum and maximum values of sequence S, while max and min are parameters that guarantee the scaled 
% sequence range falls within [-1,1]. The normalised stream sequence is shown in equation~\ref{normalized}

% \begin{equation}
%   S' = s'_1, s'_2, \dots, s'_n
%   \label{normalized}
% \end{equation}

After normalising the stream sequence to a range between -1 and 1 using Equation~\ref{minmaxscaler}, each data point $X'(t)$ is transformed into its corresponding polar coordinate angle $\phi(t)$. This transformation is accomplished by calculating the arc-cosine of $X'(t)$, as depicted in Equation~\ref{polar}. Calculating the polar angle is pivotal as it retains the numerical relationships among the sequence data points. This preservation is essential for ensuring the structural integrity of the data, which supports further analytical processes where the geometric or angular relationships between data points are crucial.

\begin{equation}
\phi(t) = \arccos(X'(t)).
\label{polar}
\end{equation}

GASF defines a specialised inner product:
$\left \langle X'(t_1), X'(t_2) \right \rangle = \cos(\phi(t_1) + \phi(t_2))$, which means that the inner product between two stream points is the cosine of the sum of the polar angles converted 
from the polar coordinates of these two stream points, in the form of:

\begin{equation}
  G=\begin{pmatrix}
      cos(\phi_1+\phi_1) & cos(\phi_1+\phi_2) & \cdots & cos(\phi_1+\phi_n)\\
      cos(\phi_2+\phi_1) & cos(\phi_2+\phi_2) & \cdots & cos(\phi_2+\phi_n)\\
      \vdots & \vdots & \cdots & \vdots \\
      cos(\phi_n+\phi_1) & cos(\phi_n+\phi_2) & \cdots & cos(\phi_n+\phi_n)
    \end{pmatrix}.
  \label{gramian}
\end{equation}

After constructing the Gram matrix, we calculate its first, second, and third moments as the features of the flow sequence as equation~\cite{gasf-ipp}.

% \begin{equation}
%   m_1=\frac{1}{N}\sum^N_{i=1}\sum^N_{j=1}G_{i,j}
%   \label{E}
% \end{equation}

% \begin{equation}
%   m_2=(\frac{1}{N}\sum^N_{i=1}\sum^N_{j=1}(G_{i,j}-\mu)^2)^{\frac{1}{2}}
%   \label{sigma}
% \end{equation}

% \begin{equation}
%   m_3 = (\frac{1}{N}\sum^N_{i=1}\sum^N_{j=1}(G_{i,j}-\mu)^3)^{\frac{1}{3}}
%   \label{M}
% \end{equation}

\begin{align}
m_1 &= \frac{1}{N} \sum_{i=1}^N \sum_{j=1}^N G_{i,j}  \\
m_2 &= \left(\frac{1}{N} \sum_{i=1}^N \sum_{j=1}^N (G_{i,j} - m_1)^2\right)^{\frac{1}{2}} \nonumber \label{gasf-ipp} \\
m_3 &= \left(\frac{1}{N} \sum_{i=1}^N \sum_{j=1}^N (G_{i,j} - m_1)^3\right)^{\frac{1}{3}} \nonumber,
\end{align}
where $N$ denotes the number of polar coordinates, and $G_{i,j}$ stands for the elements 
of matrix $G$, and $m_1$, $m_2$ and $m_3$ refer to the first, second, and third moments of matrix $G$, respectively.

Due to the capabilities of the SDN architecture, the controller is capable of aggregating global traffic data from the data plane and collecting traffic across various network protocols, including TCP and UDP streams. This functionality not only allows for the analysis of global traffic characteristics but also enables the extraction of specific features from TCP and UDP streams.

% Equations~\ref{tcp_flow} and~\ref{udp_flow} depict the TCP and UDP flows, respectively, each comprising $n$ data points.

% \begin{equation}
%   X^{TCP} = s^{TCP}_1, s^{TCP}_2, \dots, s^{TCP}_n
%   \label{tcp_flow}
% \end{equation}

% \begin{equation}
%   S^{UDP} = s^{UDP}_1, s^{UDP}_2, \dots, s^{UDP}_n
%   \label{udp_flow}
% \end{equation}

In the initial phases of a DoS-like attack, observable changes occur in the network traffic parameters: the mean of TCP flows decreases, the variance increases, and the coefficient of variation undergoes significant alterations. Due to the connectionless nature of UDP, attackers frequently exploit this protocol to inject malicious traffic into transport layer bottlenecks. As a result, we utilise the coefficient of variation of TCP streams and TCP/UDP traffic volume as discriminate features.

To summarise, the detection system captures the first, second, and third moments of the Gram matrix of global traffic, along with the coefficient of variation of TCP flows, and the number of abnormal peaks in UDP flows, to characterise network traffic. These five features are subsequently inputted into pre-trained machine learning classifiers, including GBDT, KNN, and XGBoost, to ascertain the presence of DoS-like attacks within the network.

\section{Methodology}
To enhance the adaptive capabilities of our attack method, we introduce a novel framework, termed AdaDoS, which employs adversarial reinforcement learning to design a dynamic attack strategy.

This section is structured as follows: Initially, we elaborate the threat model. Then we delineate the AdaDoS Framework, outlining the system architecture of our method. Subsequently, we delve into our proposed adversarial reinforcement learning model, detailing the decision process modelling and the training of the model. In the fourth part, we present our innovative reciprocal learning algorithm, designed to effectively address the challenges of executing attacks under constrained observational capabilities.

\subsection{Threat Model}
\noindent\textbf{Attack Scenario.}
The attack scenario revolves around a SDN environment integrated with a detection system. While the underlying network infrastructure is assumed to be trusted, malicious hosts within the network may aim to degrade the service performance of the entire network. These attackers leverage feedback from the network to bypass detection mechanisms and launch DoS attacks, exploiting the dynamic nature of SDN to degrade network quality of service.

\noindent\textbf{Attacker Capability.}
The attackers possess advanced capabilities to craft high-rate burst traffic tailored to congest the SDN. They have precise control over the duration, period, and intensity of these traffic bursts, enabling them to generate patterns that evade detection methods. The attacker can also utilize tools like the $ping$ utility to measure real-time latency along the path from the attacker’s host to the target nodes. Additionally, sufficient computational resources are accessible for attacker to deploy attack models to dynamically adjust traffic patterns in real-time based on feedback from the network environment. However, the attackers don't have any privileges of network operation.

\noindent\textbf{Attacker Prior Knowledge.}
We assume that the attackers have no knowledge of the network's topology and configuration. The attackers' understanding is restricted to the victim host's address, without insight into the broader network architecture. However, by employing tools like $iperf$, $pathload$ and $traceroute$, the attacker can estimate network bottleneck bandwidth~\cite{melander2000new}, gaining a rough understanding of the network's capacity. 

\noindent\textbf{Attack goal.}
The attackers aim to execute an efficient attack while ensuring a high survival rate. Their objectives are threefold: achieving a high attack success rate, minimizing available network bandwidth, and reducing attack costs. The attack success rate reflects the attacker’s ability to evade detection and sustain their activities over time, underscoring their survival capability. Minimizing network bandwidth measures the effectiveness of the attack in disrupting network performance, focusing on restricting the resources available for legitimate traffic. Attack cost evaluates the resources consumed by the attacker, emphasizing the need for resource-efficient strategies while maintaining maximum impact. 

\subsection{AdaDoS Framework}
The AdaDoS framework primarily comprises two components: the attacker and the detector. Both components are designed to outperform each other, forming a game-theoretic zero-sum game. Our method exploits the adversarial nature of this setup to enhance the attacker model with greater concealment and adaptability.

To enable the attacker to adapt to the characteristics of the SDN environment and the feedback from the detector, our approach employs a reinforcement learning (RL) agent. This agent interacts with both the SDN environment and the detector, updating its strategy based on their feedback.

Both limitations of observation and lack of prior knowledge is usually encountered in real-world scenarios. Inspired by recent study on Network Tomography, a technology that only relies on a few subset of network elements' information to infer internal QoS attributes, links delay within certain window-sized time is chosen as the observation of our deployed model, which can be seen as a noised rough depiction of network status \cite{he2021network}.

To further overcome the limitations of observation and improve performance, we introduce a novel reciprocal learning paradigm that utilises a pre-trained teacher model with global observation capabilities. This model facilitates policy transfer and knowledge distillation to an untrained, to-deploy student model operating under limited observational capacity. Besides, we control the student model's window size of latency measurement, making student model's input dimensionality of much lower than the teacher model's. With lower observation dimensionality and noised latency information, the deployed student attacker is expected to get more insensitive and robust to network environment variation according to \cite{gillen2021explicitly, korkmaz2024survey}.

% Network tomography has attracted significant attention as it offers the capability to reduce the monitoring overhead in a network by inferring unknown internal link states solely based on end-to-end measurements [3], [4], [5], [6], [7], [8], [9], [10], [11], [12], [21]. This technique primarily deals with additive network metrics commonly utilized in routing, such as link delay and packet loss. By selectively probing a limited set of paths from a small number of monitors positioned at the network's edge, network tomography constructs a linear system capturing the relationship between observed paths and corresponding link metrics.

\subsection{Adversarial Reinforcement Learning}
To find the solution for the formulated optimisation problem of attacker, we utilise the method of Reinforce Learning (RL), an technique that can enable model to learn from experience and future expectation.
\subsubsection{\textbf{Formulation of Markov Decision Process}}
\label{MDP}
We first formulate the process of network attack as a Markov Decision Process (MDP). With out loss of generality, the MDP of attack can be described as $<\mathcal{S}, \mathcal{A}, \mathcal{R}, \mathcal{\gamma}>$, where $\mathcal{S}$ is the state space, $\mathcal{A}$ is the action space, $\mathcal{R}$ is the reward function, and $\gamma\in (0,1)$ is a discount factor. 
The details are as follow:
\begin{enumerate}
    \item \textbf{State}\\
        The state space is
            $$\mathcal{S} \in \mathbb{R}^{m+n},$$
         where $m$ is the dimension of SDN controller's information, $n$ is the delay observation window size. 
        
        Inspired by \cite{gasf-ipp}, we use the gram-like matrix for feature extraction of traffic flow of SDN in a statistic order, and the real-time traffic volume and real-time bandwidth for the instant observation. To capture the trends of link delay and to align observation with the actual environment, the observation of attacker also contains delay of target link within given window size.
        The state can be represented as:
        \begin{equation}
            \mathbf{S} = [\textbf{s}_\text{gram}, \ \textbf{s}_\text{traffic}, \ \textbf{s}_\text{delay}]
            \label{full_s}
        \end{equation}
        \begin{itemize}
            \item $\textbf{s}_\text{gram} = [m_1, m_2, m_3]$, where $m_1, m_2, m_3$ are the first, second, and third-order moments of the gram-like matrix~\cite{gasf-ipp} of the aggregated flow during a sliding time window.
            \item $\textbf{s}_\text{traffic}=[v_\text{tcp}, v_\text{udp}, b]$, where $v_\text{tcp}$ denotes the traffic volume of TCP package, $v_\text{udp}$ denotes the traffic volume of UDP packages, and $b$ represents the available bandwidth of the target link.
            \item $\textbf{s}_\text{delay}=[\tau(t-n+1), \tau(t-n+2), \dots , \tau(t-1), \tau(t)]$ represents the delay observation of recent $n$ time slot, where $\tau(t)$ means the delay of link at time $t$, $n$ is the delay observation window size. 
        \end{itemize}
    \item \textbf{Action}\\
    The action space is 
        $$\mathcal{A} \in \{0,1\} \times \mathbb{R}^2_{\geq 0}.$$
    As a complete attack decision is incorporated with two part, when to attack and how to attack. The former part can be simplified as whether to attack in current time slot which has been discretized, and the later can be described as the attack rate and the attack duration in one time slot that both shape one attack.
    Each action $\mathbf{a}$ can be represented as:
    \begin{equation}
        \mathbf{a} = [a_\text{dec}, a_\text{rate}, a_\text{dur}],
    \end{equation}
    where $a_{dec}\ \in \{0,1\}$ is the attack decision action that decide whether to attack in current time slot, in which 0 represents not attack and 1 represents attack, $a_{rate}$ is the attack rate that defines the package sent per seconds, $a_{dur}$ is the attack duration that decides the duration of package sending. The volume of package sent in one attack duration can be calculated as:
    \begin{equation}
        v_{A} = a_{\text{dec}} \cdot a_{\text{dur}} \cdot a_{\text{rate}}.
    \end{equation}
    \item \textbf{Reward}\\
    The reward function $\mathcal{R}: \mathcal{S}\times\mathcal{A}\to\mathbb{R}$ takes input of environment state $\mathbf{S}$ and action $\mathbf{a}$, and outputs the corresponding reward, indicating the quality of the chosen action $\mathbf{a}$, which can be defined as follows:
    \begin{equation}
        \mathcal{R}(\mathbf{a}, \mathbf{S}, f_d)=
        \begin{cases}
            (z-c^2)\cdot R_c & if~ a_{\text{dec}} = 1 ~and~ f_d = 0\\
            -P & if~ a_{\text{dec}} = 1 ~and~ f_d = 1\\
            -\kappa & if~ a_{\text{dec}} = 0 ~and~ f_d = 0\\
            -P-\kappa & if~ a_{\text{dec}} = 0 ~and~ f_d = 1\\
        \end{cases},
    \end{equation}
    where $R_c$ is a constant denoting the maximum reward of an undetected attack, as known as congestion reward, $P$ is a constant meaning the penalty given when detected, $\kappa$ is the living penalty given when not attack, which is also a constant, $a_{\text{dec}}$ is the attack decision action that decide whether to attack in current time slot, $f_d$ is the detector's detection flag representing whether the detector detect the network is suffering an attack, $z$ is the clipped congestion rate of network, and $c$ is the relative cost of one attack, which is defined below.
    \begin{itemize}
        \item Congestion rate indicates the degree of network congesting, the clipped congestion rate in current time slot is defined as follows:
        \begin{equation}
            z=
            \begin{cases}
                1-\frac{b}{B_\text{max}} & if~b>B_\text{th}\\
                z_0 & if~b\leq B_\text{th}
            \end{cases},
        \end{equation}
        where $b$ is the available bandwidth of the target link, $B_\text{max}$ is the maximum bandwidth of target link, $B_\text{th}$ is the clipped threshold of bandwidth, $z_0$ is the fixed congestion rate with bandwidth under the threshold.
        \item The relative cost represent the consumed resource per attack, which is defined as follows:
        \begin{equation}
            c=\frac{a_{\text{dur}}\cdot a_{\text{rate}}}{C_\text{max}},
        \end{equation}
        where  $a_{\text{rate}}$ is the attack rate that defines the package sent per seconds, $a_{\text{dur}}$ is the attack duration that decides the duration of package sending, $C_\text{max}$ is the maximum cost per attack.
    \end{itemize}
    
    The intuition behind the reward function is to avoid detection in all conditions, which means whenever to attacker or not in current time slot, once the attacker get detected, a large penalty $P$ is given, encouraging the attacker to avert detection with full consideration.
    We also use the threshold-based reward in undetected attack reward, encouraging attacker to achieve the target congestion rate, which is represented in $z_0$ and $B_\text{th}$.
\end{enumerate}
\subsubsection{\textbf{Learning algorithm}}
With the formulated Markov Decision Process mentioned above, RL algorithm is able to applied to maximize the reward expectation and achieve the attack goal. We use the Proximal Policy Optimisation (PPO) algorithm \cite{schulman2017proximal} as the base model, which is a stable actor-critic (A-C)~\cite{Mnih2016AsynchronousMF} architecture model. In the A-C model, the actor model is used to learn the actual policy, whose output is a probability distribution, while the critic model is used to evaluate the quality of action taken by actor, in the way of generating evaluation of next state after taking action.

To better improve the attacker survival ability and attack ability, we further proposed a two-stage decision mechanism based on hierarchical reinforce learning~\cite{NEURIPS2018_HRL} considering both attack timing and attack style. We also build a deferred reward mechanism to solve the problem of feedback delay. The two-stage decision decision model is trained in an decoupled way, making the update of model accurate and efficient.

\begin{figure}[htbp]
    \centering
    \includegraphics[width=0.75\linewidth]{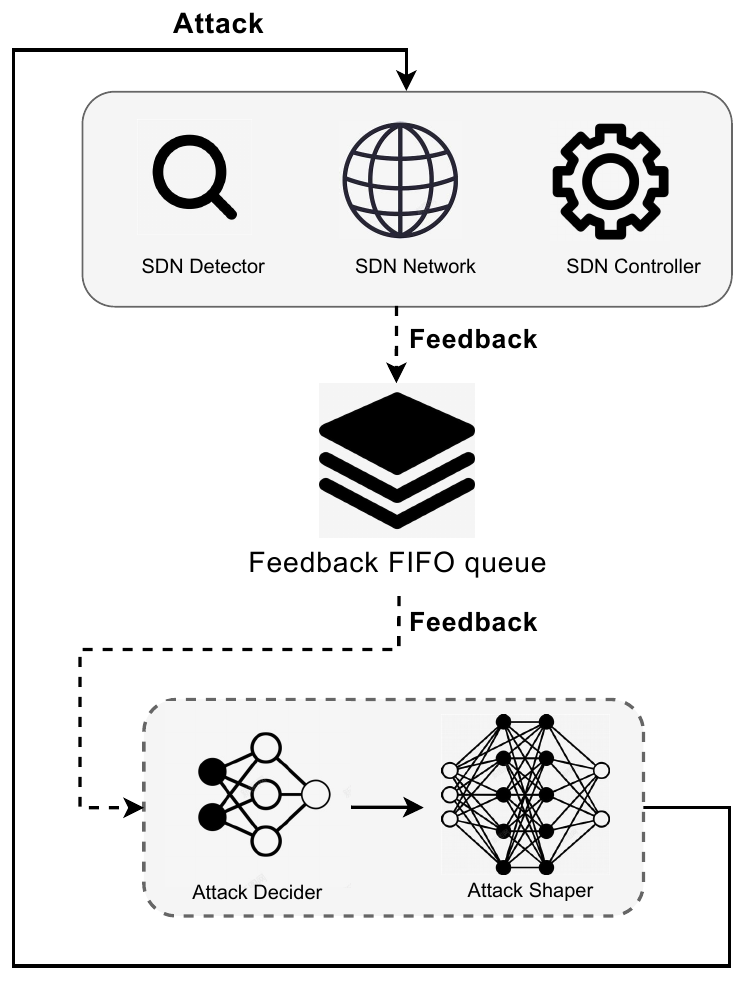}
    \caption{Adversarial RL Framework}
    \label{fig:ad_RL}
\end{figure}

\noindent\textbf{Two-stage decision mechanism. }
The attack timing and the attack style is equal important to the survival ability of attacker, since the former avoid the unsuitable timing that should not attack, and the latter improve the concealment of attack. Both of the timing choice and style choice are crucial. In some extreme examples, timing plays a critical role, such as when background traffic is low, a rushed attack is easily detected and the attack host will be blocked. Besides, the concealment built by attack style is based on the right timing choice, as the SDN traffic characteristic is usually variable in different time.
 
Considering the two sides of attack, we describe one attack action as a two-stage decision process, when to attack and how to attack. However, the decision of attack timing and the shape of attack style is in different granularity level. The attack timing choice is usually based on long-term planning, while the attack style shaping works mainly on short-term decision. To better generate policy of the long and short term policy, We model this process as a two-stage sequential process, and utilise a sequential decision model to execute this process.

 As the A-C based model is usually made up of two parts, as known as actor model and critic model, where the former generate the actual policy and the later output the evaluation of the taken action. We divide the actor model as attack \textit{decider} model and attack \textit{shaper} model. The divided two part of actor take action in a sequential style, and update in a decoupled style in their own action space. The divided model works as a hierarchical RL model, which means the action output and expectation evaluation works in a sequential order. We keep the origin architecture of critic model, which only estimate the value of state.\\
\noindent\textbf{Decoupled training process. }
As the \textit{decider} model and the \textit{shaper} model determine the timing and the manner of one attack respectively, estimates of their policies' reward expectations should differ. The \textit{decider}'s action affects all subsequent states, and therefore take responsible for all the future results. The calculate of \textit{decider}'s reward expectation should contains all the subsequent rewards. However, as attack \textit{shaper} won't take any action when \textit{decider}'s action $a_{\text{dec}}$ is $0$, the \textit{shaper}'s reward expectation only contains rewards when \textit{decider}'s action $a_{\text{dec}}$ is $1$. The Hierarchical update style is paramount for judging the quality of action, as it limits the reward expectation estimate within the scope of action, making the update of the \textit{decider} model and the \textit{shaper} model, more accurate and efficient.\\
\noindent\textbf{Deferred Reward. }
As the impact of attacks takes time to be felt by the network, the reward should be deferred. We design a first-in-first-out (FIFO) queue to store the network status and detector feedback, 
and delay the arrival of feedback to the model update module. With the reward deferred, the model is able to learn the network and detector character precisely with a more accurate feedback, which is much closer real case.

\subsection{Reciprocal Learning}
The real-world practice of AdaDoS usually meets the problem of lacking network information. We formulate this problem as a partially observable Markov Decision Process (POMDP) problem, and leverage the Mutual-learned Teacher-student scheme to solve the lack of observation.

\begin{figure*}[htbp]
    \centering
    \includegraphics[width = 0.95\linewidth,angle=-90,scale=0.4]{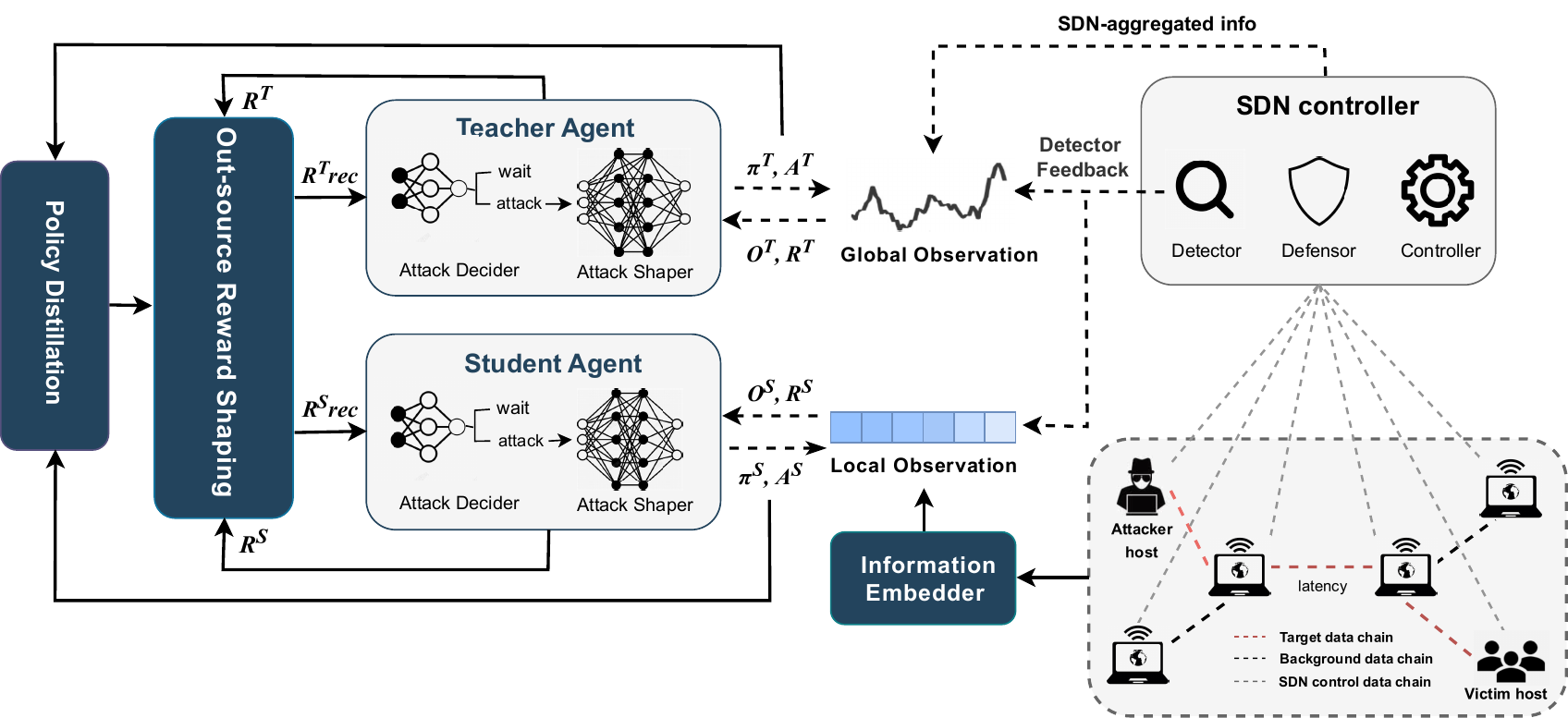}
    \caption{Reciprocal Learning}
    \label{fig:Ts-RL}
\end{figure*}
\subsubsection{Formulating the POMDP Problem}
 The process of network attack under limited observation is formulated as POMDP. Without loss of generality, the POMDP of attack can be described as a 5-tuple, $<\mathcal{S}, \mathcal{A}, \mathcal{O},\mathcal{R}, \mathcal{\gamma}>$, where $\mathcal{O}$ is the observation space, $\mathcal{S}$, $\mathcal{A}$, $\mathcal{R}$ is the the state space, action space and reward function defined before in \ref{MDP}, $\gamma\in (0,1)$ is a discount factor in MDP definition.
Considering the Extensibility of our model, our definition of the attacker agent's observation only contains links delay, which is obtained by attacker agent itself with tool like $ping$ etc. 
The observation space under partial observation is
        $$\mathcal{O} \in \mathbb{R}^{n'},$$
where $n'$ is the delay observation window size under limited observation circumstances. 
The attacker's observation $\textbf{\textit{O}}$ at current time $t$ can be defined as:
\begin{equation}
    \textbf{\textit{O}}=[\tau(t-n'+1), \tau(t-n'+2), \dots , \tau(t-1), \tau(t)]
\end{equation}
where $\tau(t)$ means the delay of link $l$ at time $t$, $n'$ is the delay observation window size under limited-observation circumstances, which is usually much smaller than teacher's delay observation window size $n$.

\subsubsection{Policy Transfer}
Finding an optimised policy under limited observation is typically challenging. To address this issue, we utilise knowledge from a pre-trained attacker agent that possesses full observation, as discussed in Section~\ref{MDP}. This pre-trained agent, endowed with complete observation, serves as the \textit{teacher} agent, while the agent being trained with restricted observation assumes the role of the \textit{student}. Our objective is to facilitate the transfer of policy from the teacher agent to the student agent. As the policy is a conditional distribution of observation with parameters, the transfer of policy is equal to the transfer a probability distribution. 
% The process shown by formula \ref{ts_update_para} is equal to the process described below:
% \begin{equation}
% \label{ts_update_para}
%     \theta_{s'} \gets \theta_s - \lambda\cdot g(\pi_s^{\theta_s}, \pi_t^{\theta_t}, r_s, r_t, \textbf{S}_t, \textbf{\textit{O}}_s)
% \end{equation}
% where $\pi_t^{\theta_t}$ is the policy of teacher agent and $\pi_s^{\theta_s}$ denotes the policy of student agent. $\theta_s$ is the parameter of student policy model $\pi_s$, $\theta_t$ is the parameter of teacher policy model $\pi_t$, $r_s$ is the set of student reward of all the time, $r_t$ is the set of teacher reward of all the time, $g$ is the policy transfer function with the input of two policies and two reward sets, and $\lambda$ is a parameter adjusting the transfer rate.

Inspired by \cite{KL_RL} and \cite{rs_RL} and based on reinforcement learning update mechanism, we can consolidate the transfer update on parameter into the rectification of reward and use the method of reward shaping to conduct the transfer process. The policy transfer with reward rectification can be formulated as follows:
\begin{equation}
\label{ts_update_rwd}
    r_S^{rec} = r_S - \lambda_{r}\cdot g(\pi_S^{\theta_S}, \pi_T^{\theta_T}, r_S, r_T, \textbf{\textit{O}}_T, \textbf{\textit{O}}_S),
\end{equation}
where $\pi_T^{\theta_T}$ denotes the policy of teacher agent and $\pi_S^{\theta_S}$ denotes the policy of student agent, $\textbf{\textit{O}}_T$ is the observation of teacher agent, which is equivalent to the $\textbf{S}$ defined in equation~\ref{full_s}, $\textbf{\textit{O}}_S$ is the observation of the student agent, $r_S$ is the student reward, $r_S$ is the set of teacher reward of all the time, $r_S^{rec}$ is the rectified reward, $g$ is the reward-based policy transfer function with the input of two policies and two reward, which is shown by the line 12 to 18 in algorithm \ref{alg_ts}, the $\lambda_{r}$ is a hyper-parameter adjusting the transfer rate.

The utilisation of the reward rectification greatly simplify the training process. It also leverage the update module of RL algorithm, improving the stability and efficiency of transfer learning process, contributing to better performance of student model.
 
\subsubsection{Reciprocal Learning Mechanism}
The policy transfer depends on a stable and robust teacher model, whose policy will be distilled to the student model. However, due to the variable environment of SDN, the robustness and consistency of the pre-trained teacher are not guaranteed. We propose the reciprocal learning mechanism to mitigate this problem. The teacher and student model is trained in an mutual learning style under reciprocal learning mechanism. To specify the mechanism in the view  of reward shaping, it means that when the teacher's reward is higher than the student's reward, only the student's reward will be rectified, but when student's reward is the higher one, both of the student and teacher's rewards will be rectified. With such a mutual learning paradigm, both of the teacher and student will improve compared to the model before, and the high-quality policy outputted by teacher model will in turn enhance the training of student model. The details of reward rectification is shown in algorithm~\ref{alg_ts}.

\noindent\textbf{Policy Kullback-Leibler Divergence. }The A-C model is usually based on stochastic policy model, which is hard to measure diffent policy with few concrete action. To measure difference of the teacher's and student's policy, we introduce the Policy Kullback-Leibler (KL) Divergence, which can directly measure the variance between two distribution, as well as stochastic policy. To better estimate the policy KL divergence in training, We first treat the policy as the joint distribution of state and action, and then convert it as a conditional distribution based on state, which is the actual format of the model outputted policy. We also rewrite the integration process in origin KL divergence as the expectation of the sampling of policy. The policy KL divergence between teacher and student policy is described as follows:
\begin{equation}
    \begin{aligned}
        \mathcal{D}_{\mathrm{KL}}\left(\pi_{T}||\pi_{S}\right)& =\int_{\textbf{\textit{O}}_T,A_T}\pi_T(\textbf{\textit{O}}_T,A_T)\cdot\ln\left(\frac{\pi_T(\textbf{\textit{O}}_T,A_T)}{\pi_S(\textbf{\textit{O}}_T,A_T)}\right) \\
        &=E_{(\textbf{\textit{O}}_T,A_t)\sim\pi_T}\left[\ln\left(\frac{\pi_T(A_T|\textbf{\textit{O}}_T)}{\pi_S(A_T|\textbf{\textit{O}}_T)}\right)\right],
    \end{aligned}
\end{equation}
where $\pi_S$ and $\pi_T$ are the distribution format policy of teacher and student, $A_T$ is the action taken by teacher, $\mathcal{S}_T$ indicates the policy-related teacher state set.

The rewrote policy KL divergence allows us to directly measure the difference of two policy outputted by two different model through policy sampling, and the process of estimation can be conducted during the training process, which proves practical in training. 

\noindent\textbf{Adaptive Fine-tune.} The robustness and consistency of the pre-trained teacher are also related to the feedback of the SDN environment. To make the teacher model better adapted to a new SDN environment, we propose the Adaptive Fine-tune for the teacher model update. As is shown in algorithm \ref{alg_ts}, the parameter $k$ is the coefficient that defined the teacher model update rate of environment. With the adjustment of environment effect on teacher model update, pre-trained model will be able to keep performance and output high-quality policy continuously for the student model. 
% \noindent\textbf{Transfer Rate Decay. } As the \\

\section{Experiments and Evaluation}

\subsection{Experimental Setup}

We conducted extensive experiments using a network simulation platform to assess the performance of AdaDoS. The hardware configuration for these experiments included an Intel Core i5-13600K CPU with 32GB of RAM and an NVIDIA GeForce RTX 4070 GPU equipped with 12GB of graphics memory. The experiments were performed on a software platform running Ubuntu 22.04.

\subsubsection{SDN Configurations and Hyper-parameter Settings}
In line with most research papers on DoS-like attack and defence, our experimental setup utilises a Ryu controller to emulate the control plane of a Software-Defined Network (SDN) and employs Mininet to simulate the corresponding data plane. This approach, widely recognised and adopted in SDN security research~\cite{Revathi2022, mohsin2022performance, alwabisi2022using}, is the standard in the field. Utilising the OpenFlow protocol, the Ryu controller periodically issues event requests to extract traffic information from data layer switches, encompassing global, TCP, and UDP traffic metrics. In our tests, the Ryu controller was configured to obtain traffic data from the data layer at a minimum interval of 0.5 seconds. We leveraged this rapid cycle to collect real-time network traffic data, which was instrumental in constructing a comprehensive dataset for our detection system. We utilise $iperf$ for real-time bandwidth assessments on bottleneck links. $Iperf$, a widely recognised tool, measures TCP and UDP bandwidth performance dynamically. Additionally, we employ Ping to evaluate link latency and to approximate network congestion levels. Ping, a ubiquitous network diagnostic tool, operates by sending ICMP echo requests to measure the round-trip time (RTT) and packet loss across various computers, functioning autonomously without the need for user interaction or specialised configuration on the target system.

Table~\ref{Hyperparameter} presents the key hyperparameters employed in the experiment.
\begin{table}[htbp]
\footnotesize
\caption{Hyper-Parameter Settings}
\label{Hyperparameter}
    \setlength{\tabcolsep}{10mm}{\begin{tabular}{ll}
        \hline
        Parameter Setting              & Value       \\
        \hline
        Episode simulation time (s)        & 100         \\
        Controller sample interval (s)     & 0.5         \\
        Decision time slot (s)             & 1           \\
        Reward delay time (s)              & 0.5         \\
        Discount rate $\gamma$             & 0.95        \\
        Clipped congestion rate $z_0$      & 1           \\
        Bandwidth thresh $B_\text{th}$ (Mbps)   & 0.1*$B_\text{max}$ \\
        Living penalty $\kappa$            & 5           \\
        Detected penalty $p$               & 100         \\
        Congestion reward $R_c$            & 80         \\
        \hline
    \end{tabular}}
\end{table}

\subsubsection{Dataset}
This paper replays the WIDE dataset as background traffic for experiments. The WIDE dataset is a traffic data repository maintained by the MAWI working group of the WIDE project in Japan~\cite{MAWIHomePage}. This traffic data repository collects real traffic data from the backbone network and provides traffic tracking data from seven different sampling points over different time periods (hours to tens of hours) since 2006. These traffic tracking data are collected by the Tcpdump tool and include protocols such as TCP, UDP, and ICMP. Based on the traffic data collected at sampling point F on January 1, 2018 from the WIDE dataset~\cite{MAWI2018}, we processed and generated a Pacp format file. Specifically, we filtered out packets other than the transport layer protocol in the original dataset, retaining only data using TCP, UDP, and ICMP protocols, and modified the source and destination IP addresses of the packets using the Scapy library.

% \subsubsection{Baselines}

\subsubsection{Metric}
Four performance metrics are recorded as LDoS attack's parameters and AdaDoS attack's strategy details, including \textbf{duration}, \textbf{period}, \textbf{attack rate} and \textbf{trigger rate}. The duration means the duration of package sending per attack. The attack rate defines the package sent per seconds in current attack. The period decides the frequency to launch an attack. The trigger rate represents the average possibility for attacker to launch an attack.

We also use two evaluation metrics to evaluate the AdaDoS attack performance, namely, \textbf{attack success rate} and \textbf{available bandwidth}. Both metrics are the average values for each episode. The higher the attack success rate, the better the concealment of the attack; The lower available bandwidth, the higher the network congestion caused by the attack. Additionally, we utilise \textbf{attack cost} to measure the cost of the attack. The attack success rate is calculated by equation~\ref{atk_success_rate} and bandwidth is measured by $iperf$. For LDoS, attack cost refers to the total number of transmitted data per attack cycle. For AdaDoS, attack cost is the average total number of transmitted data per episode.

\begin{equation}
    \text{Attack success rate} = \frac{\text{No. undetected attacks}}{\text{No. attacks}}.
    \label{atk_success_rate}
\end{equation}

Note that the attack success rate and available bandwidth are the two most significant indicators of attack performance, as the primary objective of all SDN attack algorithms is to maximise effectiveness. Although attack cost is also a consideration for our algorithm, it is of lower priority. In real-world practice, acquiring additional resources is typically easier than achieving a higher success rate.

\subsection{Attack Performance}
\subsubsection{The Attack Capability of AdaDoS} \label{simp_exp}
To verify the attack capability of AdaDoS, we conducted 10 groups of LDoS attacks with different parameters, and compared their attack success rate and bandwidth with AdaDoS. Each group of LDoS attacks lasts 10000 seconds to ensure that we get enough experimental data.
We use LDoS as the baseline for two main reasons: (1) No other work has shown significantly better performance than LDoS in DoS-like attack area, and (2) most recent, high-quality research papers~\cite{gasf-ipp, tang2021performance} use LDoS as the baseline for both attack and defence purposes. Given the critical and well-recognised threat posed by LDoS attacks, our comprehensive analysis within this context is sufficient to validate the effectiveness of AdaDoS. As such, comparisons with a broader range of attack types are not necessary for establishing the primary contributions of our work.

\begin{figure}[htbp]
    \centering
    \includegraphics[width=0.45\textwidth]{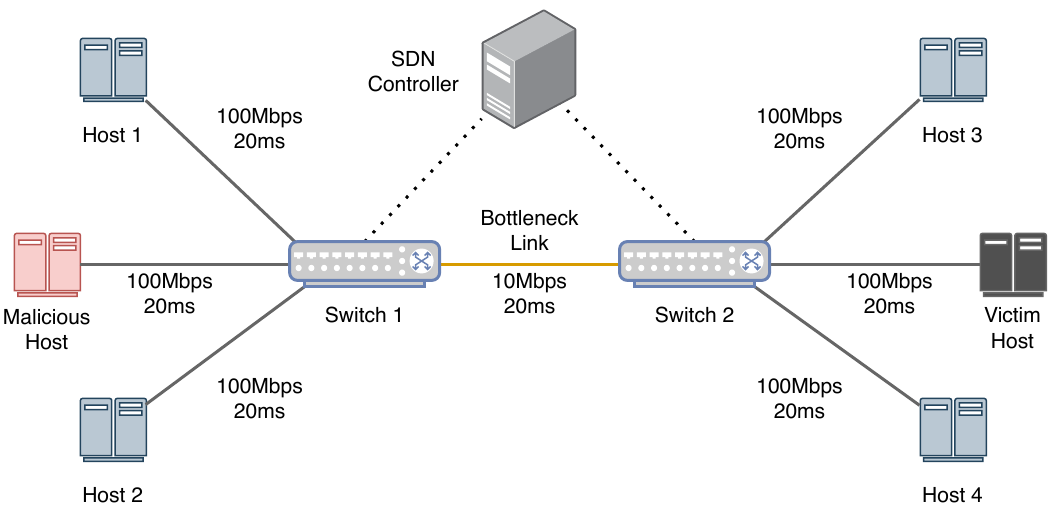}
    \caption{Simple Topology}
    \label{fig:simple_topo}
\end{figure}

According to most recent related works~\cite{gasf-ipp, tang2021performance}, we use a simple topology as the network topology, as shown in Figure~\ref{fig:simple_topo}. Among them, host 1 sends the WIDE dataset background traffic to host 3, and the malicious host sends the attack traffic to the victim host. Host 2 and host 4 use $iperf$ to measure the real-time TCP bandwidth of bottleneck links.

Table~\ref{table:simple_adados} presents the performance comparison between LDoS and AdaDoS attack. Compared to LDoS, AdaDoS not only achieves a higher attack success rate but also significantly increases network congestion. When detected by detector, each group of LDoS attacks had a very low attack success rate, whereas AdaDoS skilfully evades the detection and achieves higher bandwidth congestion. Although our algorithm incurs a higher attack cost compared to the baseline, this trade-off is justified in the context of real-world applications. In practice, the availability of additional resources often outweighs the challenge of increasing the attack success rate. Furthermore, it is important to note that our algorithm's cost is substantially lower than that of a normal DoS attack. Thus, the enhanced effectiveness and efficiency of our algorithm make it a superior choice despite the higher cost, as the ultimate goal is to achieve more reliable and impactful attacks while maintaining a reasonable resource expenditure.

\subsubsection{AdaDoS Robustness with Various SDN Detectors}

To verify the robustness of AdaDoS attacks under various detectors, we deployed different detectors to detect AdaDoS attacks. With the global traffic monitoring capabilities of the SDN controller, we can collect and analyse the data of attack traffic, label the data, and input it into the detector for training. Based on the experimental data of LDoS attacks, we can train an LDoS detector. Based on the traffic data of AdaDoS attacks, we can train an AdaDoS detector to detect AdaDoS attacks under existing strategies. Finally, we mix the LDoS attack data and AdaDoS attack data, or use a mix detector with more comprehensive detection capabilities. This group of experiments is conducted on a simple topology. Similarly, host 1 sends WIDE dataset background traffic to host 3, and malicious host sends attack traffic to the victim host. Host 2 uses $iperf$ tool to measure the TCP bandwidth of bottleneck links between host 2 and host 4.

\begin{figure}[htbp]
    \centering
    \includegraphics[width=0.45\textwidth]{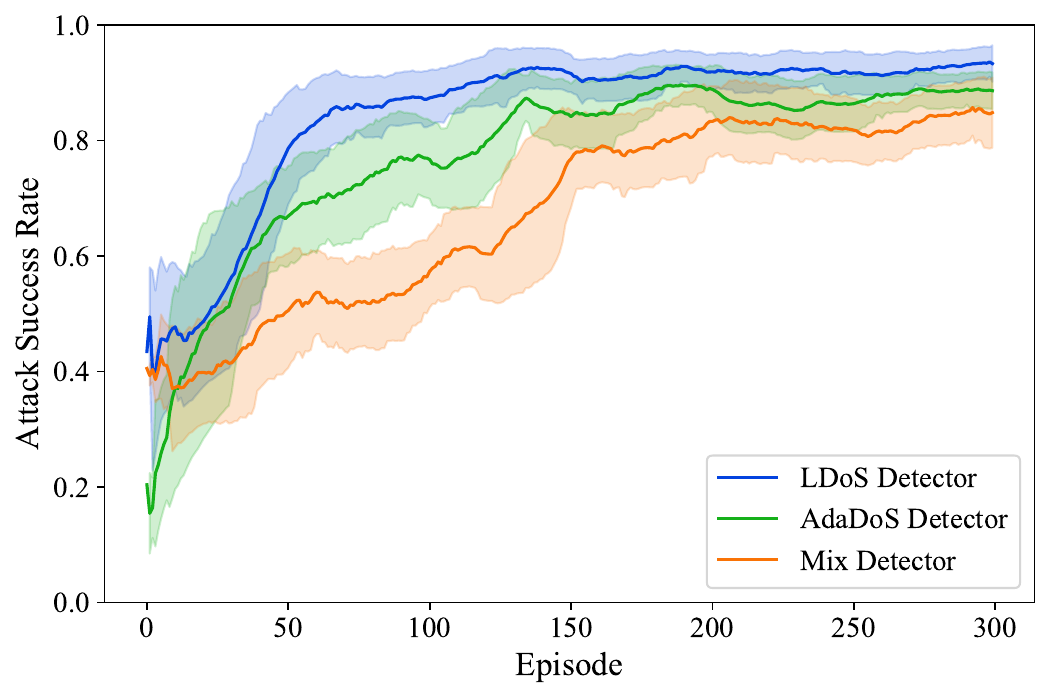}
    \caption{AdaDoS Against Different Detectors}
    \label{fig:mix_detector}
\end{figure}

Figure~\ref{fig:mix_detector} displays the curves for the attack success rates of the AdaDoS attack across three different detectors. It is evident that the AdaDoS attack ultimately achieves a higher success rate under each detector. Initially, a notable observation is the lower success rate of the AdaDoS attack against the AdaDoS detector, which is specifically trained with data from previous AdaDoS attacks. Despite this initial challenge, the AdaDoS attack demonstrates remarkable adaptability; subsequent training sessions significantly enhance its attack strategies. This evolution underscores the AdaDoS attack's sophisticated capability to continually refine and evolve its methods in response to defensive measures.

% To validate the robustness of AdaDoS against complex network topology in real-world scenarios, we conducted experiments using the network topology provided by Aarnet~\cite{AARNet2024}, as depicted in Figure~\ref{fig:aarnet}. Hosts 1 and 2 deliver background traffic from the WIDE dataset to the victim host, while host 3 measures real-time TCP bandwidth with Iperf. The malicious Host delivers attack traffic to the Victim host, aiming to congest the link between them. In Aarnet, we executed three sets of LDoS attack experiments using various parameters as baselines, and each group of experiments also lasted 10000 seconds.

% Table~\ref{table:aarnet} presents the attack performance for LDoS and AdaDoS in the Aarnet topology. AdaDoS outperforms LDoS in terms of attack success rate while maintaining a significantly higher bandwidth blocking rate. This indicates that AdaDoS can adapt to diverse network topology and adjust its strategies to achieve optimal attack outcomes. 

% \begin{figure}[htbp]
%     \centering
%     \includegraphics[width=\linewidth]{Aarnet.pdf}
%     \caption{Aarnet Topology}
%     \label{fig:aarnet}
% \end{figure}

\subsubsection{AdaDoS Robustness with Various Network Topologies}
To validate the robustness of AdaDoS against complex network topologies in real-world scenarios, we conducted experiments using three network topologies: Aarnet, Ansnet, and Yorknet. Among three topologies, the Aarnet and Yorknet topologies are large-scale network topologies, while the Ansnet is a medium-scale network topology. The selection of these three topologies is intended to demonstrate the scalability of our method with respect to network topologies with different scale and complexity, and to provide a comprehensive comparison of the performance of LDoS and AdaDoS in different network environments. We respectively executed three sets of LDoS attack experiments in these topologies using various parameters as baselines, and each group of experiments also lasted 10000 seconds. All detailed of these topologies are shown in Appendix.

We directly deployed the attacker model trained on the simple topology shown in Figure~\ref{fig:simple_topo} on these three complex networks. Table~\ref{table:aarnet} presents the attack performance for LDoS and AdaDoS in various topologies. AdaDoS outperforms LDoS in terms of attack success rate while maintaining a significantly higher bandwidth congestion rate in all three topologies. The trigger rate varies in topologies with different complexity. This indicates that AdaDoS can adapt to diverse network topology and adjust its strategies to achieve optimal attack outcomes. 

\subsection{Reciprocal Learning Performance}
\begin{figure*}[ht]
\centering
\begin{subfigure}[b]{0.3\textwidth}
    \includegraphics[width=\textwidth]{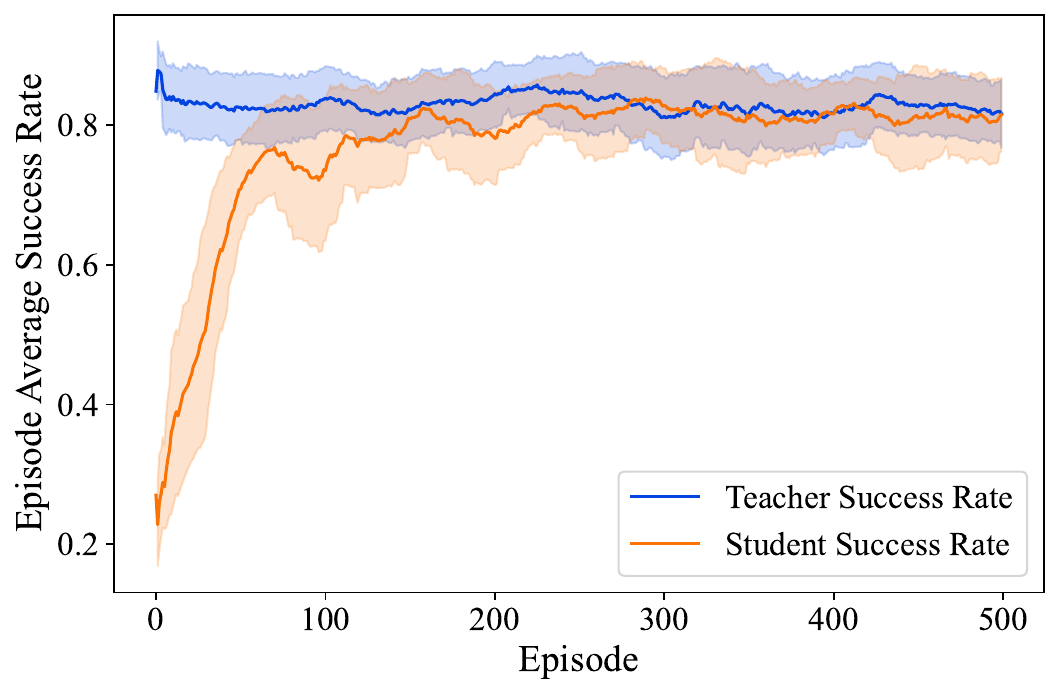}
    \caption{Attack Success Rate}
    \label{fig:ts_succ}
\end{subfigure}
\hfill
\begin{subfigure}[b]{0.3\textwidth}
    \includegraphics[width=\textwidth]{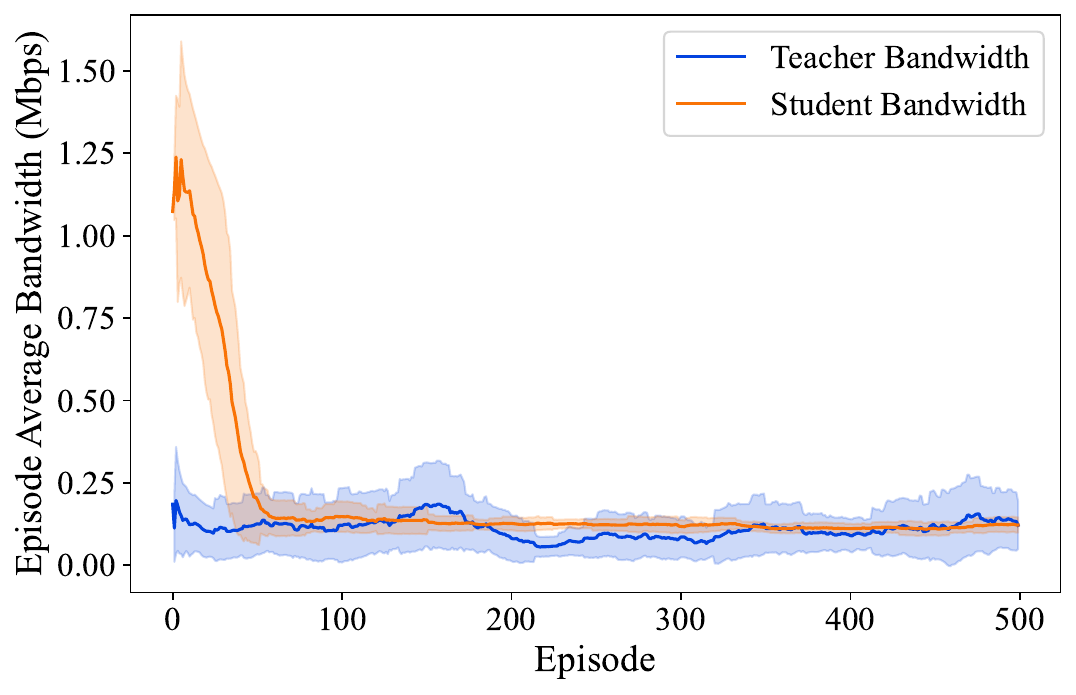}
    \caption{Available Bandwidth}
    \label{fig:ts_bw}
\end{subfigure}
\hfill
\begin{subfigure}[b]{0.3\textwidth}
    \includegraphics[width=\textwidth]{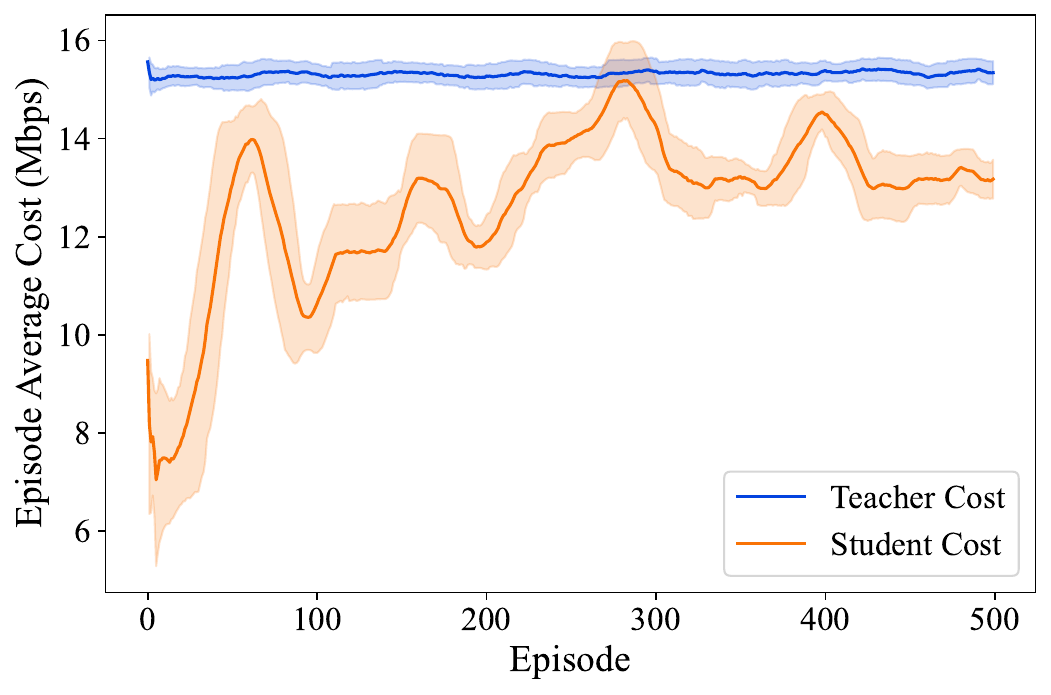}
    \caption{Attack Cost}
    \label{fig:ts_cost}
\end{subfigure}
\caption{Reciprocal Learning Performance}
\label{fig:ts_comparision}
\end{figure*}

To validate the effectiveness of the reciprocal learning mechanism, we test our reciprocal learning framework performance on the simple topology with WIDE dataset as background traffic, and compare the teacher and student performance within 500 episodes.

Figure~\ref{fig:ts_comparision} illustrates the training curves of teacher and student model on three metrics. The teacher model is pre-trained and updated only with reciprocal learning mechanism, while the student model is updated with both reciprocal learning and environment feedback. As shown by the figure \ref{fig:ts_comparision}, the student model achieve about the same attack success rate and bandwidth congestion compared to the teacher, and even outperform the teacher model on the metric of attack average cost. The experiment demonstrates the potential of our reciprocal learning framework, which enable the student attacker with limited observation to reach the similar attack performance of teacher model with a complete observation.

\subsection{Noise Disturbance Influence}
\begin{figure}[htbp]
    \centering
    \includegraphics[width=0.49\textwidth]{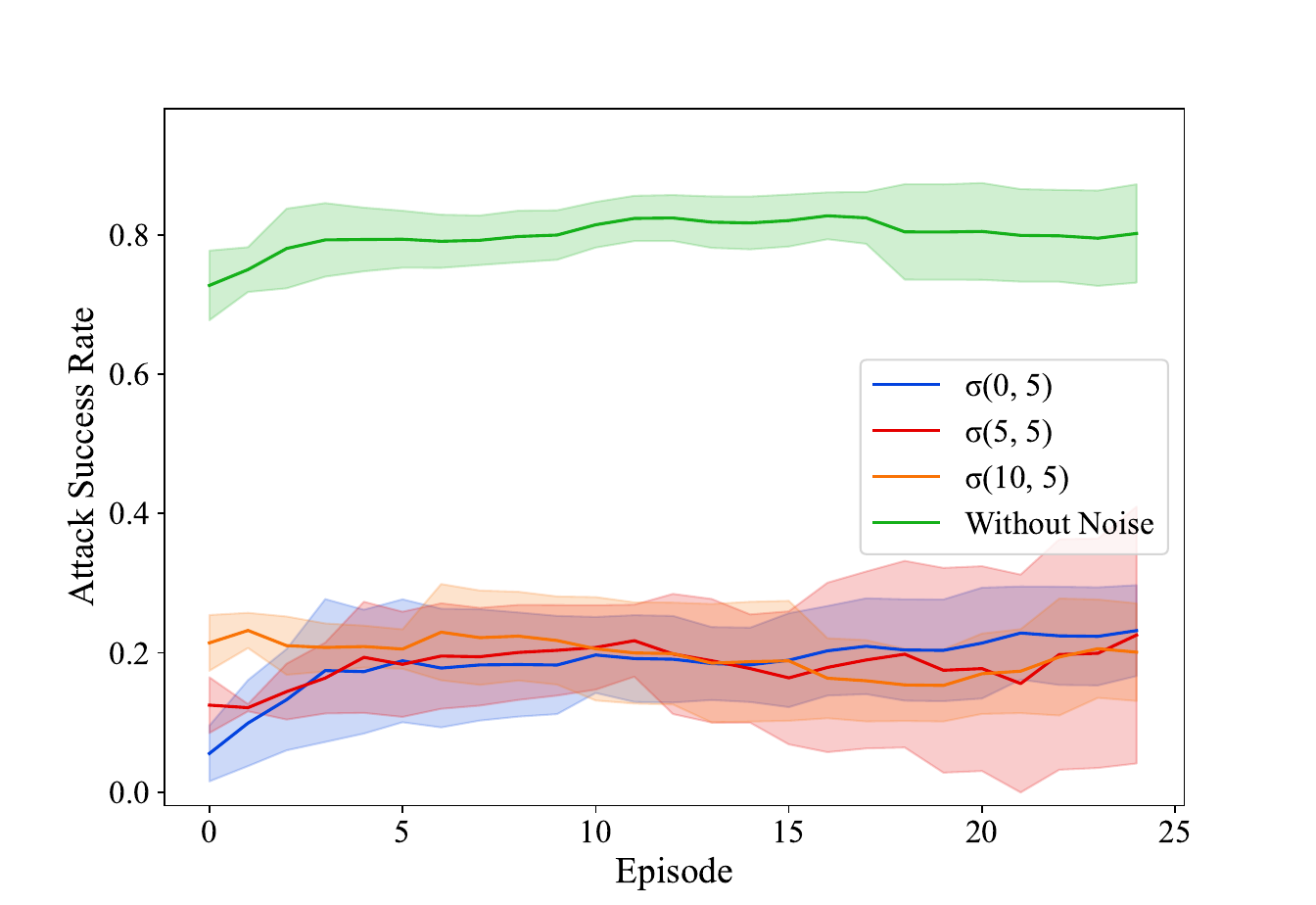}
    \caption{Noise Disturbance on AdaDoS}
    \label{fig:noise_ada}
\end{figure}

As our method relies on observed delay information, introducing noise can significantly impact the performance of AdaDoS. To assess the noise sensitivity of AdaDoS, we added three sets of Gaussian noise to the delay information observed by the agent in the environment of a simple topology, maintaining the same settings as in Section \ref{simp_exp}. We then compared the attack performance with and without noise.  Fig~\ref{fig:noise_ada} displays the attack success rate (ASR) of AdaDoS under three sets of Gaussian noise. The depicted curves show the trained attack agent's ASR under various levels of delay observation noise. The substantial reduction in ASR between experiments with and without noise highlights AdaDoS's vulnerability to delay information disturbance. This finding suggests that adding noise to delay information can be an effective countermeasure against AdaDoS.

Despite this vulnerability, our attack approach remains valuable for demonstrating the importance of accurate delay information in SDN environments and for developing more robust SDN defence mechanisms. It is important to note, however, that adding noise to delay information may have side effects, such as degrading the overall network performance and potentially affecting legitimate traffic. Therefore, while noise addition can serve as a countermeasure, it must be carefully calibrated to mitigate these adverse effects. By understanding AdaDoS's sensitivity to noise, researchers can better appreciate the trade-offs involved in SDN security and work towards more comprehensive solutions that balance attack effectiveness and defence strategies.

\begin{table*}[!ht]
\footnotesize
\caption{The attack capability of AdaDoS and the LDoS baselines with different parameters. For the attack success rate and the average bandwidth index, the mean value is indicated on the left of the sign, while the standard deviation is shown on the right. '/' indicates the value is not applicable.}
\label{table:simple_adados}
\begin{tabular}{cclllllll}
\hline
\multirow{2}{*}{No.} & \multirow{2}{*}{Type}  & \multirow{2}{*}{Duration (s)} & \multirow{2}{*}{Period (s)} & \multirow{2}{*}{Attack Rate (Mbps)} & \multirow{2}{*}{\begin{tabular}[c]{@{}l@{}}Trigger rate\end{tabular}} & \multirow{2}{*}{\begin{tabular}[c]{@{}l@{}}Attack Success \\ Rate\end{tabular}} & \multirow{2}{*}{\begin{tabular}[c]{@{}l@{}}Average Bandwidth\\ (Mbps)\end{tabular}} & \multirow{2}{*}{\begin{tabular}[c]{@{}l@{}}Attack Cost \\ (Mbps)\end{tabular}}  \\
                     &                        &                               &                             &                                     &                                                                                 &                                                                                     &                                                                                \\ \hline
1                    & \multirow{10}{*}{LDoS} & 0.15                          & 1.0                         & 15              & 1.00                     & $0.254\pm0.039$                                                                    & $5.3\pm0.044$                                                                           & 2.25                                                                          \\
2                    &                        & 0.15                          & 1.5                         & 15             & 1.00                     & $0.257\pm0.044$                                                                    & $5.4\pm0.122$                                                                           & 1.50                                                                           \\
3                    &                        & 0.20                          & 1.5                         & 15             & 1.00                     & $0.262\pm0.044$                                                                    & $5.4\pm0.126$                                                                           & 2.00                                                                           \\
4                    &                        & 0.15                          & 1.0                         & 20             & 1.00                     & $0.237\pm0.036$                                                                    & $5.3\pm0.195$                                                                           & 3.00                                                                           \\
5                    &                        & 0.15                          & 1.5                         & 20             & 1.00                     & $0.255\pm0.044$                                                                    & $5.4\pm0.148$                                                                           & 2.00                                                                           \\
6                    &                        & 0.20                          & 1.5                         & 20             & 1.00                     & $0.269\pm0.045$                                                                    & $5.4\pm0.134$                                                                           & 2.67                                                                           \\
7                    &                        & 0.15                          & 2.0                         & 25             & 1.00                     & $0.234\pm0.050$                                                                    & $5.6\pm0.176$                                                                           & 1.88                                                                           \\
8                    &                        & 0.15                          & 2.0                         & 30             & 1.00                     & $0.235\pm0.049$                                                                    & $5.6\pm0.172$                                                                           & 2.25                                                                           \\
9                    &                        & 0.10                          & 2.0                         & 30             & 1.00                     & $0.225\pm0.045$                                                                    & $5.6\pm0.134$                                                                           & 1.50                                                                           \\ \hline
10 & \multirow{10}{*}{\begin{tabular}[c]{@{}l@{}}Double-\\LDoS\end{tabular}}& 0.15  & 1.0    & 15  & 1.00  & $0.150\pm0.447$  & $0.3\pm0.056$   & 6.63      \\
    11 &                               & 0.15  & 1.5    & 20 & 1.00  & $0.270\pm0.054$  & $0.1\pm0.242$   & 5.81      \\
    12 &                               & 0.20  & 1.5    & 20 & 1.00  & $0.282\pm0.052$  & $0.1\pm0.367$   & 8.39      \\
    13 &                               & 0.15  & 1.0    & 20 & 1.00  & $0.302\pm0.074$  & $0.1\pm0.313$   & 7.17      \\
    14 &                               & 0.15  & 1.5    & 15 & 1.00  & $0.278\pm0.077$  & $0.1\pm0.236$   & 7.64      \\
    15 &                               & 0.20  & 1.5    & 15 & 1.00  & $0.285\pm0.069$  & $0.1\pm0.212$   & 8.68      \\
    16 &                               & 0.20  & 1.5    & 25 & 1.00  & $0.311\pm0.063$  & $0.2\pm0.326$   & 23.24      \\
    17 &                               & 0.15  & 2.0    & 30 & 1.00  & $0.314\pm0.064$  & $0.1\pm0.243$   & 10.57     \\
    18 &                               & 0.10  & 2.0    & 30 & 1.00  & $0.294\pm0.078$  & $0.2\pm0.371$   & 26.89     \\
19                   &                        & Rand(0.1, 0.15, 0.2)          & Rand(1.0, 1.5, 2.0)         & Rand(15, 20, 25, 30)           & 1.00     & $0.254\pm0.004$    & $5.2\pm0.105$    & /    \\ \hline
20                   & \bf{AdaDoS}          & $0.55\pm0.21$                             & 1.0                           & $9.00\pm3.92$             & $0.95\pm0.02$                      & $\bf{0.795}\pm{0.062}$                                                           & $\bf{0.1}\pm{0.126}$                                                                  & 4.78                                                                           \\ \hline
\end{tabular}
\end{table*}

% \begin{table}
%   \caption{Frequency of Special Characters}
%   \label{tab:freq}
%   \begin{tabular}{ccl}
%     \toprule
%     Non-English or Math&Frequency&Comments\\
%     \midrule
%     \O & 1 in 1,000& For Swedish names\\
%     $\pi$ & 1 in 5& Common in math\\
%     \$ & 4 in 5 & Used in business\\
%     $\Psi^2_1$ & 1 in 40,000& Unexplained usage\\
%   \bottomrule
% \end{tabular}
% \end{table}

\begin{table*}[htbp]
    \centering
    \small
    \caption{The robustness of AdaDoS in three extensive topology and the LDoS baseline with different parameters. For the adaDoS rows, the mean value is indicated on the left of the sign, while the standard deviation is shown on the right. '/' indicates the value is not applicable.}
    \label{table:aarnet}
    \setlength{\tabcolsep}{1.1mm}{\begin{tabular}{lcllllllll}
    \hline
    No. & Topology & Type   & Duration (s) & Period (s) & Attack Rate (Mbps) & \begin{tabular}[c]{@{}l@{}}Trigger Rate\end{tabular} & \begin{tabular}[c]{@{}l@{}}Attack Success \\ Rate\end{tabular} & \begin{tabular}[c]{@{}l@{}}Average Bandwidth\\ (Mbps)\end{tabular} & \begin{tabular}[c]{@{}l@{}}Attack Cost\\ (Mbps)\end{tabular} \\ \hline
1  &         &        & 0.15         & 1.0        & 232.5     & 1.00         & $0.370\pm0.051$      & $99.9\pm1.55$          & 34.91            \\
2  &  & LDoS   & 0.20         & 1.5        & 310.0      & 1.00        & $0.372\pm0.057$      & $100.4\pm1.66$         & 41.33                  \\
3  & Aarnet        &        & 0.10         & 2.0        & 465.0    & 1.00          & $0.364\pm0.062$      & $100.5\pm2.05$         & 23.32                  \\
4  &         & \bf{AdaDoS} & $0.46\pm0.19$            & 1.0         & $229.5\pm72.6$    & $0.61\pm0.09$    & $\bf{0.79}\pm{0.091}$     & $\bf{53.31}\pm{24.95}$   & 63.08                  \\
\hline
5  &         &        & 0.15         & 1.0        & 75.0     & 1.00      & $0.268\pm0.063$      & $3.0\pm51.13$          & 11.34                    \\
6  &  &  LDoS  & 0.20         & 1.5        & 100.0       & 1.00       & $0.426\pm0.103$      & $0.6\pm0.94$         & 13.31                            \\
7  & Ansnet        &        & 0.10         & 2.0        & 150.0     & 1.00      & $0.295\pm0.123$      & $0.4\pm0.96$         & 7.50                            \\
8  &         & \bf{AdaDoS} & $0.25\pm0.24$            & 1.0          & $76.2\pm18.5$    & $0.59\pm0.05$    & $\bf{0.871}\pm{0.062}$     & $\bf{1.129}\pm{0.54}$   & $22.51$                \\
\hline
9  &         &        & 0.15         & 1.0        & 225.0    & 1.00          & $0.560\pm0.071$      & $71.8\pm1.72$          & 33.81                  \\
10  &  & LDoS   & 0.20         & 1.5        & 300.0      & 1.00        & $0.605\pm0.075$      & $72.3\pm2.12$         & 40.04                   \\
11  & Yorknet        &        & 0.10         & 2.0        & 450.0      & 1.00        & $0.603\pm0.072$      & $71.6\pm2.62$         & 22.52                   \\
12  &         & \bf{AdaDoS} & $0.50\pm0.21$            & 1.0          & $216.8\pm63.5$   & $0.21\pm0.09$     & $\bf{0.795}\pm{0.118}$     & $\bf{26.32}\pm{9.74}$   & 59.08        \\ 
\hline
\end{tabular}}
\end{table*}

\section{Defence scheme}
As previously mentioned, existing rule-based or machine learning-based detectors struggle to identify AdaDoS, as it adapts its attack patterns in response to detector results. The adaptive capabilities of AdaDoS necessitate innovative defence strategies beyond traditional methods. In this section, we discuss possible countermeasures that network administrators can be used to mitigate the attack.

\noindent\textbf{Disturbing Attacker Observaion.}
One inherent limitation of AdaDoS is its reliance on delay information to orchestrate attacks and it's vulnerability in observation noise, as shown in Fig \ref{fig:noise_ada}. This dependency offers a strategic point of vulnerability that can be exploited to mitigate the threat. A straightforward defence approach involves deliberately introducing noise into the delay data. This can be achieved by occasionally altering routing paths intentionally or by withholding packages to randomly inject noise into delay information between nodes. Although these tactics can significantly impair the ability of AdaDoS to execute precise attacks, they are not without their drawbacks. Introducing noise or withholding information can disrupt the normal operations of the network, potentially degrading service quality and affecting legitimate network traffic. Therefore, while these strategies offer a viable means of defence against AdaDoS, they also necessitate a careful consideration of the trade-offs between security and network performance.

\noindent\textbf{Adopting Zero-trust Architecture.}
Adopting a zero-trust architecture is recommended as a fundamental security enhancement. This architecture requires continuous authentication and authorisation of all users, regardless of their position relative to the network perimeter, ensuring rigorous security checks and continuous validation of security configurations before granting or maintaining access to data and network resources. However, overly strict Zero-Trust defenses may inadvertently block innocent traffic; thus, mechanisms like adaptive thresholds, traffic verification feedback loops, and a quarantine mode for further analysis should be integrated to minimize false positives and maintain service availability.

\noindent\textbf{Leveraging Adversarial Trained Detector.}
A suggested way to boost SDN defense capability is integrating adversarial training into the network’s defence mechanisms. This involves training models to recognise and mitigate inputs designed using adversarial attack techniques, potentially enhancing the network's resilience against strategies similar to those used by AdaDoS. However, there is scepticism regarding the effectiveness of this approach. Experiences with generative adversarial networks (GANs)~\cite{Goodfellow2020} have shown that distinguishing between genuine and AI-generated images has become increasingly difficult. This raises concerns about the adaptability of adversarial training, as such models might struggle to differentiate between legitimate network activities and those mimicking AdaDoS patterns.

\section{Conclusion}
This paper introduce AdaDoS, an adaptive adversarial reinforcement learning-based attack framework designed to generate DoS-like malware under partial observation SDN environment. To address the low survival ability problem, AdaDoS employs an innovative deep reinforcement learning model and propose a two-stages structure to dynamically change the attack patterns. To address the limited observation problem from attacker side, AdaDoS propose a reciprocal learning framework with teacher-student structure. Furthermore, we evaluate the performance of AdaDoS against latest detectors via real-world settings, finding that it can bypass these defences. The findings underscore the potential of AdaDoS in the field of attacks in SDN and possible any SDN-like network.

\newpage

\section*{Ethics Considerations}
In developing AdaDoS, an advanced adversarial reinforcement learning-based framework for simulating DoS-like attacks in SDN environments, we acknowledge the dual-use nature of this research and the ethical challenges it poses. Our primary objective is to enhance understanding of network vulnerabilities and strengthen defenses against increasingly sophisticated attacks. However, we recognise the potential for misuse of the framework and are committed to addressing this concern responsibly by adhering to ethical principles outlined in relevant resources, including the USENIX Security guidelines.

To mitigate risks, all details regarding the tactics, techniques, and procedures (TTPs) employed in AdaDoS will be shared only with verified entities in the academic and security research communities. Such entities must undergo a rigorous vetting process, including verification of credentials and a formal commitment to ethical use, ensuring that the framework supports defensive efforts without contributing to offensive capabilities. We also emphasise the importance of weighing harms and benefits under both consequentialist and deontological ethics, striving to prevent tangible harms and respect individuals’ rights, as recommended by the principles of Beneficence and Respect for Persons.

Furthermore, all experimental activities are conducted exclusively in controlled and isolated environments designed to emulate real-world network configurations without connecting to operational systems. These environments employ virtualized topologies and synthetic traffic while incorporating real-time monitoring and automated safeguards to detect and prevent any anomalous or unintended behaviour. This ensures that research does not inadvertently disrupt real-world systems or compromise the security of external networks.

In addition, we proactively consider the full spectrum of stakeholders potentially impacted by this research, including network operators, end users, and the broader cybersecurity community. By addressing ethical considerations at every stage of the research process and documenting these decisions transparently, we aim to align with global best practices for ethical cybersecurity research. In cases where ethical analyses under different principles lead to conflicting conclusions, we commit to articulating and justifying our decisions clearly, as recommended by USENIX Security guidelines.

Finally, we adhere to all applicable laws and regulations governing cybersecurity research, including disclosure practices for vulnerabilities. Vulnerabilities discovered during this research are disclosed promptly and responsibly to affected parties, in alignment with the principle of minimizing harm and maximizing positive outcomes. This rigorous approach ensures that AdaDoS contributes to advancing the field of network security while upholding the highest ethical standards. 
% during the paper review process to prevent any potential misuse. 

\section*{Compliance with the Open Science Policy}
We will open the source of AdaDoS through GitHub at this link (\url{https://anonymous.4open.science/r/AdaDoS}).

\bibliographystyle{plain}
\bibliography{\jobname}

\begin{thebibliography}{10}

\bibitem{AARNet2024}
{AARNet}.
\newblock Australia's academic and research network, 2024.
\newblock Date of visit: 2024-04-29.

\bibitem{ahmad2015security}
Ijaz Ahmad, Suneth Namal, Mika Ylianttila, and Andrei Gurtov.
\newblock Security in software defined networks: A survey.
\newblock {\em IEEE Communications Surveys \& Tutorials}, 17(4):2317--2346, 2015.

\bibitem{alwabisi2022using}
Sulaiman Alwabisi, Ridha Ouni, and Kashif Saleem.
\newblock Using machine learning and software-defined networking to detect and mitigate ddos attacks in fiber-optic networks.
\newblock {\em Electronics}, 11(23):4065, 2022.

\bibitem{rs_RL}
Tim Brys, Anna Harutyunyan, Matthew~E Taylor, and Ann Nowé.
\newblock Policy transfer using reward shaping.
\newblock In {\em Autonomous Agents and Multiagent Systems}, pages 181--188, 2015.

\bibitem{cao2019crosspath}
Jiahao Cao, Qi~Li, Renjie Xie, Kun Sun, Guofei Gu, Mingwei Xu, and Yuan Yang.
\newblock The {CrossPath} attack: Disrupting the {SDN} control channel via shared links.
\newblock In {\em 28th USENIX Security Symposium (USENIX Security 19)}, pages 19--36, Santa Clara, CA, August 2019. USENIX Association.

\bibitem{cao2018disrupting}
Jiahao Cao, Mingwei Xu, Qi~Li, Kun Sun, Yuan Yang, and Jing Zheng.
\newblock Disrupting sdn via the data plane: A low-rate flow table overflow attack.
\newblock In Xiaodong Lin, Ali Ghorbani, Kui Ren, Sencun Zhu, and Aiqing Zhang, editors, {\em Security and Privacy in Communication Networks}, pages 356--376, Cham, 2018. Springer International Publishing.

\bibitem{dongping16}
Ping Dong, Xiaojiang Du, Hongke Zhang, and Tong Xu.
\newblock A detection method for a novel ddos attack against sdn controllers by vast new low-traffic flows.
\newblock In {\em 2016 IEEE International Conference on Communications (ICC)}, pages 1--6, 2016.

\bibitem{ezekiel2017dynamic}
Sharon Ezekiel, Dinil~Mon Divakaran, and Mohan Gurusamy.
\newblock Dynamic attack mitigation using sdn.
\newblock In {\em 2017 27th International Telecommunication Networks and Applications Conference (ITNAC)}, pages 1--6, 2017.

\bibitem{gillen2021explicitly}
Sean Gillen and Katie Byl.
\newblock Explicitly encouraging low fractional dimensional trajectories via reinforcement learning.
\newblock In {\em Conference on Robot Learning}, pages 2137--2147. PMLR, 2021.

\bibitem{giotis2016scalable}
Kostas Giotis, George Androulidakis, and Vasilis Maglaris.
\newblock A scalable anomaly detection and mitigation architecture for legacy networks via an openflow middlebox.
\newblock {\em Security and Communication Networks}, 9(13):1958--1970, 2016.

\bibitem{Goodfellow2020}
Ian Goodfellow, Jean Pouget-Abadie, Mehdi Mirza, Bing Xu, David Warde-Farley, Sherjil Ozair, Aaron Courville, and Yoshua Bengio.
\newblock Generative adversarial networks.
\newblock {\em Communications of the ACM}, 63:139--144, 2020.

\bibitem{2017STAR}
Zehua Guo, Ruoyan Liu, Yang Xu, Andrey Gushchin, Anwar Walid, and H.~Jonathan Chao.
\newblock Star: Preventing flow-table overflow in software-defined networks.
\newblock {\em Computer Networks}, 125(oct.9):15--25, 2017.

\bibitem{he2021network}
Ting He, Liang Ma, Ananthram Swami, and Don Towsley.
\newblock {\em Network tomography: identifiability, measurement design, and network state inference}.
\newblock Cambridge University Press, 2021.

\bibitem{hsu2015design}
Shihwen Hsu, Tsengyi Chen, Yungchun Chang, Shuohan Chen, Hanchieh Chao, Tsenyeh Lin, and Weikuan Shih.
\newblock Design a hash-based control mechanism in vswitch for software-defined networking environment.
\newblock In {\em 2015 {IEEE} International Conference on Cluster Computing}, pages 498--499, Chicago, IL, USA, 2015. {IEEE} Computer Society.

\bibitem{hu2015comprehensive}
Zhiyuan Hu, Mingwen Wang, Xueqiang Yan, Yueming Yin, and Zhigang Luo.
\newblock A comprehensive security architecture for sdn.
\newblock In {\em 2015 18th International Conference on Intelligence in Next Generation Networks}, pages 30--37, 2015.

\bibitem{topo_zoo}
Simon Knight, Hung~X. Nguyen, Nickolas Falkner, Rhys Bowden, and Matthew Roughan.
\newblock The internet topology zoo.
\newblock {\em IEEE Journal on Selected Areas in Communications}, 29(9):1765--1775, 2011.

\bibitem{korkmaz2024survey}
Ezgi Korkmaz.
\newblock A survey analyzing generalization in deep reinforcement learning.
\newblock {\em arXiv preprint arXiv:2401.02349}, 2024.

\bibitem{li2021ldos}
Xinmeng Li, Kai Zheng, Dan Tang, Zheng Qin, Zhiqing Zheng, and Shihan Zhang.
\newblock Ldos attack detection based on asnnc-ofa algorithm.
\newblock In {\em 2021 IEEE Wireless Communications and Networking Conference (WCNC)}, pages 1--6, 2021.

\bibitem{melander2000new}
Bob Melander, Mats Bjorkman, and Per Gunningberg.
\newblock A new end-to-end probing and analysis method for estimating bandwidth bottlenecks.
\newblock In {\em Globecom'00-IEEE. Global Telecommunications Conference. Conference Record (Cat. No. 00CH37137)}, volume~1, pages 415--420. IEEE, 2000.

\bibitem{meti2017detection}
Nisharani Meti, D~G Narayan, and V.~P. Baligar.
\newblock Detection of distributed denial of service attacks using machine learning algorithms in software defined networks.
\newblock In {\em 2017 International Conference on Advances in Computing, Communications and Informatics (ICACCI)}, pages 1366--1371, 2017.

\bibitem{Mnih2016AsynchronousMF}
Volodymyr Mnih, Adri{\`a}~Puigdom{\`e}nech Badia, Mehdi Mirza, Alex Graves, Timothy~P. Lillicrap, Tim Harley, David Silver, and Koray Kavukcuoglu.
\newblock Asynchronous methods for deep reinforcement learning.
\newblock In {\em International Conference on Machine Learning}, 2016.

\bibitem{mohsin2022performance}
Mayadah~A Mohsin and Ali~H Hamad.
\newblock Performance evaluation of sdn ddos attack detection and mitigation based random forest and k-nearest neighbors machine learning algorithms.
\newblock {\em Revue d'Intelligence Artificielle}, 36(2):233, 2022.

\bibitem{NEURIPS2018_HRL}
Ofir Nachum, Shixiang~(Shane) Gu, Honglak Lee, and Sergey Levine.
\newblock Data-efficient hierarchical reinforcement learning.
\newblock In S.~Bengio, H.~Wallach, H.~Larochelle, K.~Grauman, N.~Cesa-Bianchi, and R.~Garnett, editors, {\em Advances in Neural Information Processing Systems}, volume~31. Curran Associates, Inc., 2018.

\bibitem{pascoal2020slow}
Tulio~A Pascoal, Iguatemi~E Fonseca, and Vivek Nigam.
\newblock Slow denial-of-service attacks on software defined networks.
\newblock {\em Computer Networks}, 173:107223, 2020.

\bibitem{ScikitLearn}
Fabian Pedregosa.
\newblock Scikit-learn: Machine learning in python, 2011.

\bibitem{phan2016novel}
Trung~V. Phan, Nguyen~Khac Bao, and Minho Park.
\newblock A novel hybrid flow-based handler with ddos attacks in software-defined networking.
\newblock In {\em 2016 Intl IEEE Conferences on Ubiquitous Intelligence \& Computing, Advanced and Trusted Computing, Scalable Computing and Communications, Cloud and Big Data Computing, Internet of People, and Smart World Congress (UIC/ATC/ScalCom/CBDCom/IoP/SmartWorld)}, pages 350--357, 2016.

\bibitem{prakash2018intelligent}
Aditya Prakash and Rojalina Priyadarshini.
\newblock An intelligent software defined network controller for preventing distributed denial of service attack.
\newblock In {\em 2018 Second International Conference on Inventive Communication and Computational Technologies (ICICCT)}, pages 585--589. IEEE, 2018.

\bibitem{Revathi2022}
M~Revathi, V~V Ramalingam, and B~Amutha.
\newblock A machine learning based detection and mitigation of the ddos attack by using sdn controller framework.
\newblock {\em Wireless Personal Communications}, pages 1--25, 2022.

\bibitem{sattar2016adaptive}
Danish Sattar, Ashraf Matrawy, and Olufemi Adeojo.
\newblock Adaptive bubble burst (abb): Mitigating ddos attacks in software-defined networks.
\newblock In {\em 2016 17th International Telecommunications Network Strategy and Planning Symposium (Networks)}, pages 50--55. IEEE, 2016.

\bibitem{schulman2017proximal}
John Schulman, Filip Wolski, Prafulla Dhariwal, Alec Radford, and Oleg Klimov.
\newblock Proximal policy optimization algorithms.
\newblock {\em arXiv preprint arXiv:1707.06347}, 2017.

\bibitem{shang2017flooddefender}
Gao Shang, Peng Zhe, Xiao Bin, Hu~Aiqun, and Ren Kui.
\newblock Flooddefender: Protecting data and control plane resources under sdn-aimed dos attacks.
\newblock In {\em IEEE INFOCOM 2017-IEEE Conference on Computer Communications}, pages 1--9. IEEE, 2017.

\bibitem{shirali2013efficient}
Sajad Shirali-Shahreza and Yashar Ganjali.
\newblock Efficient implementation of security applications in openflow controller with flexam.
\newblock In {\em 2013 IEEE 21st annual symposium on high-performance interconnects}, pages 49--54. IEEE, 2013.

\bibitem{tang2022new}
Dan Tang, Jingwen Chen, Xiyin Wang, Siqi Zhang, and Yudong Yan.
\newblock A new detection method for ldos attacks based on data mining.
\newblock {\em Future Generation Computer Systems}, 128:73--87, 2022.

\bibitem{tang2023detection}
Dan Tang, Chenjun Gao, Xinmeng Li, Wei Liang, Sheng Xiao, and Qiuwei Yang.
\newblock A detection and mitigation scheme of ldos attacks via sdn based on the fss-rsr algorithm.
\newblock {\em IEEE Transactions on Network Science and Engineering}, 2023.

\bibitem{gasf-ipp}
Dan Tang, Siyuan Wang, Boru Liu, Wenqiang Jin, and Jiliang Zhang.
\newblock Gasf-ipp: Detection and mitigation of ldos attack in sdn.
\newblock {\em IEEE Transactions on Services Computing}, 16(5):3373--3384, 2023.

\bibitem{tang2021performance}
Dan Tang, Yudong Yan, Siqi Zhang, Jingwen Chen, and Zheng Qin.
\newblock Performance and features: Mitigating the low-rate tcp-targeted dos attack via sdn.
\newblock {\em IEEE Journal on Selected Areas in Communications}, 40(1):428--444, 2021.

\bibitem{thomas2017ddos}
Roshni~Mary Thomas and Divya James.
\newblock Ddos detection and denial using third party application in sdn.
\newblock In {\em 2017 International Conference on Energy, Communication, Data Analytics and Soft Computing (ICECDS)}, pages 3892--3897. IEEE, 2017.

\bibitem{wang2015floodguard}
Haopei Wang, Lei Xu, and Guofei Gu.
\newblock Floodguard: A dos attack prevention extension in software-defined networks.
\newblock In {\em 2015 45th Annual IEEE/IFIP International Conference on Dependable Systems and Networks}, pages 239--250. IEEE, 2015.

\bibitem{Wang2015/12}
Mingxin Wang, Huachun Zhou, Jia Chen, and Bo~Tong.
\newblock An approach for protecting the openflow switch from the saturation attack.
\newblock In {\em Proceedings of the 2015 4th National Conference on Electrical, Electronics and Computer Engineering}, pages 729--734. Atlantis Press, 2015/12.

\bibitem{wang2015imaging}
Zhiguang Wang and Tim Oates.
\newblock Imaging time-series to improve classification and imputation, 2015.

\bibitem{MAWIHomePage}
{WIDE Project}.
\newblock Mawi working group traffic archive.
\newblock \url{https://mawi.wide.ad.jp/mawi/}.

\bibitem{MAWI2018}
{WIDE Project}.
\newblock Samplepoint-f 2018-01-01 14:00 traffic data.
\newblock \url{https://mawi.wide.ad.jp/mawi/samplepoint-F/2018/201801011400.html}, 2018.

\bibitem{wu2019sequence}
Zhijun Wu, Qingbo Pan, Meng Yue, and Liang Liu.
\newblock Sequence alignment detection of tcp-targeted synchronous low-rate dos attacks.
\newblock {\em Computer Networks}, 152:64--77, 2019.

\bibitem{5696753}
Yang Xiang, Ke~Li, and Wanlei Zhou.
\newblock Low-rate ddos attacks detection and traceback by using new information metrics.
\newblock {\em IEEE Transactions on Information Forensics and Security}, 6(2):426--437, 2011.

\bibitem{xie2020research}
Shengxu Xie, Changyou Xing, Guomin Zhang, Xianglin Wei, and Guyu Hu.
\newblock Research on ldos attack detection and defense mechanism in software defined networks.
\newblock In {\em Security and Privacy in Social Networks and Big Data: 6th International Symposium, SocialSec 2020, Tianjin, China, September 26--27, 2020, Proceedings 6}, pages 85--96. Springer, 2020.

\bibitem{KL_RL}
Haiyan Yin and Sinno~Jialin Pan.
\newblock Knowledge transfer for deep reinforcement learning with hierarchical experience replay.
\newblock In {\em Proceedings of the Thirty-First AAAI Conference on Artificial Intelligence}, AAAI'17, page 1640–1646. AAAI Press, 2017.

\bibitem{2019Defending}
Bin Yuan, Deqing Zou, Shui Yu, Hai Jin, Weizhong Qiang, and Jinan Shen.
\newblock Defending against flow table overloading attack in software-defined networks.
\newblock {\em Services Computing, IEEE Transactions on}, 12(2):231--246, 2019.

\bibitem{2018Joint}
Gongming Zhao, Hongli Xu, Shigang Chen, Liusheng Huang, and Pengzhan Wang.
\newblock Joint optimization of flow table and group table for default paths in sdns.
\newblock {\em IEEE/ACM Transactions on Networking}, PP(4):1--14, 2018.

\end{thebibliography}

% \newpage

\section*{Appendix}
\subsubsection*{Network Topology}\label{appendix_topology}
\begin{figure}[!ht]
\centering
\begin{subfigure}[b]{0.45\textwidth}
    \includegraphics[width=\textwidth]{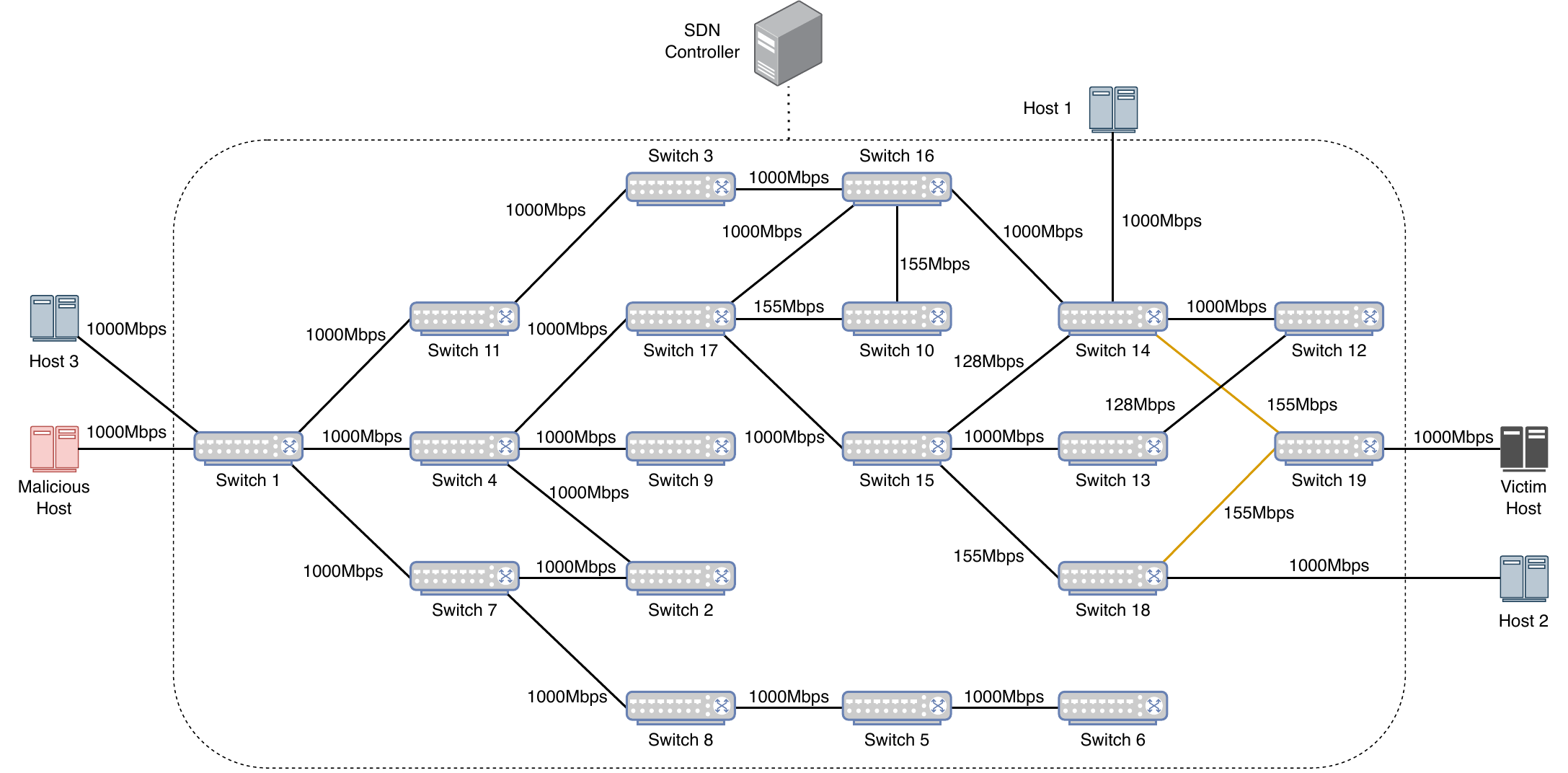}
    \caption{Aarnet}
    \label{fig:aarnet}
\end{subfigure}
\hfill
\begin{subfigure}[b]{0.45\textwidth}
    \includegraphics[width=\textwidth]{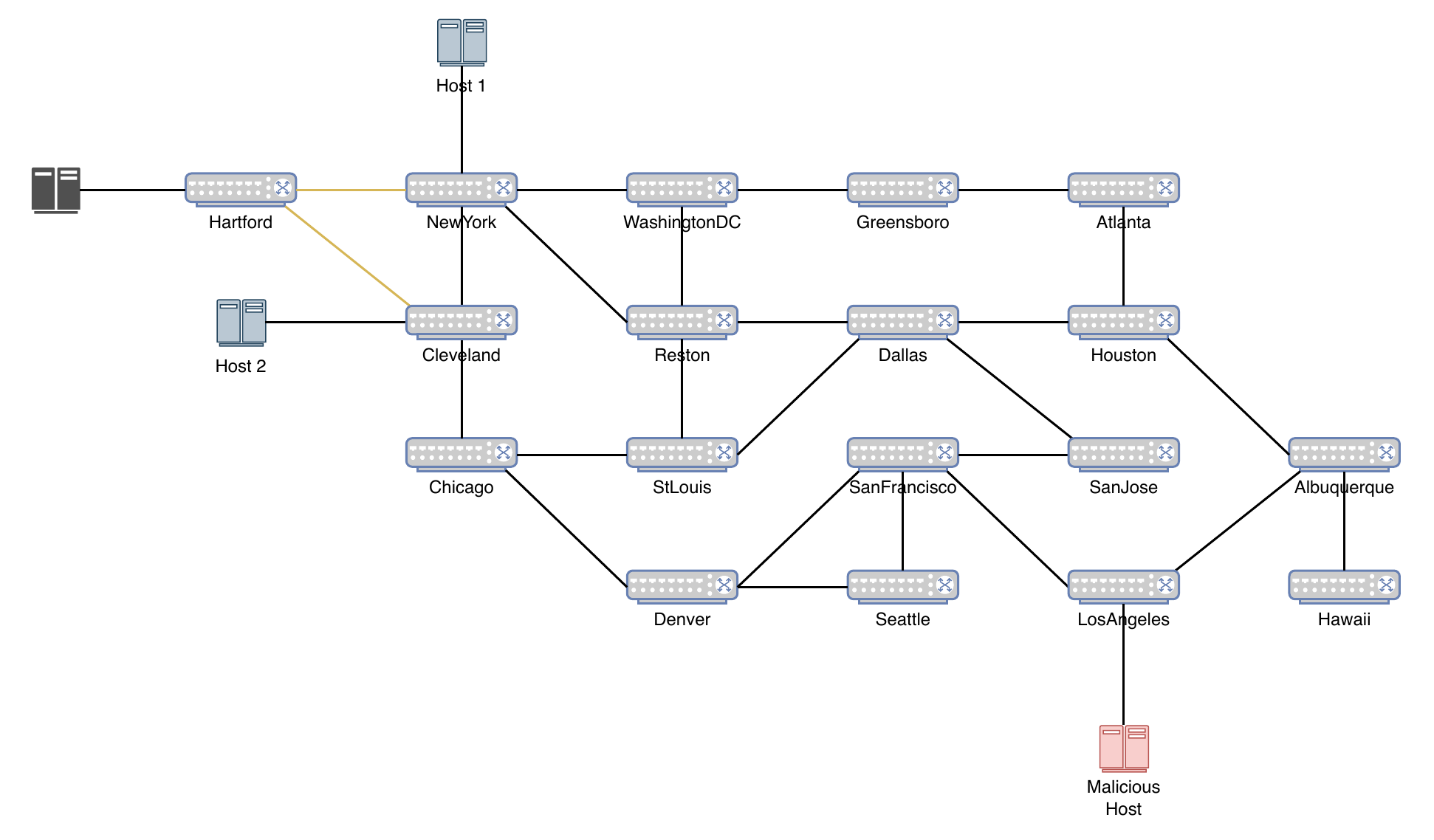}
    \caption{Ansnet}
    \label{fig:topo_ans}
\end{subfigure}
\hfill
\begin{subfigure}[b]{0.45\textwidth}
    \includegraphics[width=\textwidth]{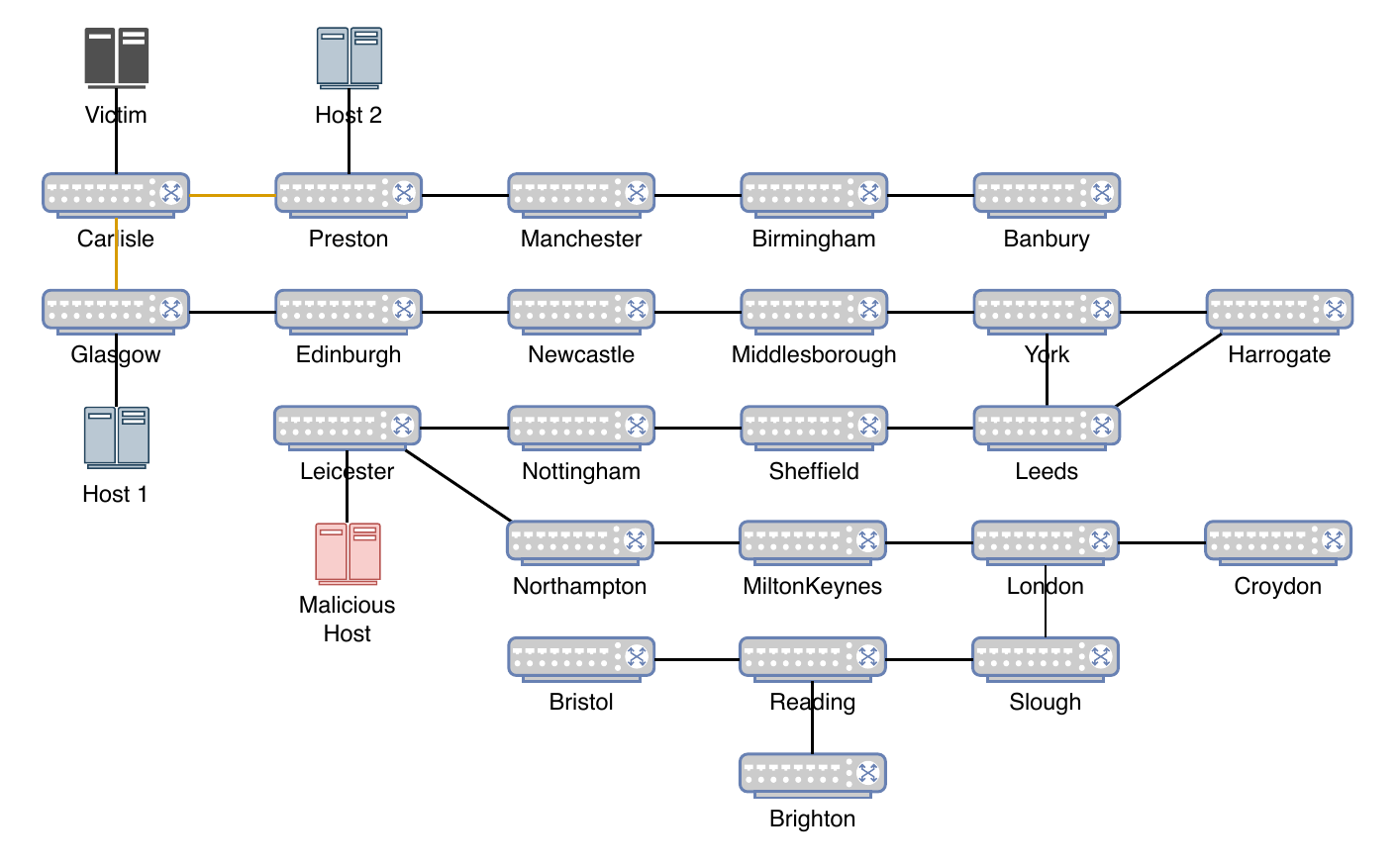}
    \caption{Yorknet}
    \label{fig:topo_york}
\end{subfigure}
\caption{Extra Network Topologies}
\label{fig:all_topo}
\end{figure}

Aarnet is provided by~\cite{AARNet2024}, as depicted in Figure~\ref{fig:aarnet}. Hosts 1 and 2 deliver background traffic from the WIDE dataset to the victim host, while host 3 measures real-time TCP bandwidth with $iperf$. The malicious host delivers attack traffic to the victim host, aiming to congest the link between them. Similar to Aarnet, background traffic from dataset are replayed on the Ansnet and Yorknet from the Internet Topology Zoo~\cite{topo_zoo}, whose topologies are shown in Figure~\ref{fig:topo_ans} and Figure~\ref{fig:topo_york}, and experiments are conducted on them in the same way.

\subsubsection*{The Algorithm of Reciprocal Learning}
The Reciprocal Learning algorithm leverages a collaborative learning paradigm between a student and a teacher agent. Both agents operate in their respective environments, \(E_S\) for the student and \(E_T\) for the teacher. The teacher begins with a pre-trained actor-critic model, while the student learns from scratch. In each iteration, the agents take actions based on their respective policies, observe new states, and receive rewards. The rewards are adjusted via reciprocal learning mechanisms, incorporating a KL-divergence-based regularization to encourage policy alignment. The teacher aids the student by scaling and modifying the reward function based on performance differences. Both agents update their actor-critic networks using temporal-difference (TD) errors derived from the adjusted rewards with Generalized Advantage Estimation (GAE). This process iteratively improves the student’s performance while maintaining collaboration between the agents.
\begin{algorithm}
\caption{Reciprocal Learning} 
\label{alg_ts}
\begin{algorithmic} 
\STATE {\bf Input:} Environment of student $E_S$ and Environment of teacher $E_T$, learning rate of student and teacher critic model $\alpha_S, \alpha_T$, learning rate of student and teacher actor model $\beta_S, \beta_T\in[0,1]$, transfer rate of student and teacher $\lambda_{S}, \lambda_T$, reward scale rate $k\in [0,1]$, discount rate $\gamma\in [0,1]$,  Pre-trained teacher actor-critic network $\pi_T^{\theta_T}$ ,$V_T^{\omega_T}$
\STATE {\bf Output:} actor-critic network $\pi_S^{\theta_S}, V_S^{\omega_S}$
\STATE \textbf{Initialise} environment $E_S, E_T$ and parameters of student actor-critic network $\pi_S^{\theta_S}, V_S^{\omega_S}: \theta_S, \omega_S\in\mathbb{R}^h$
\FOR{each iteration}
\STATE observe $\textbf{\textit{O}}_S, \textbf{\textit{O}}_T$
\WHILE{state not terminal}
    \STATE $a_S\sim \pi_S (\cdot|\textbf{\textit{O}}_S;~\theta_S)$
    \STATE $a_T\sim \pi_T (\cdot|\textbf{\textit{O}}_T;~\theta_T)$
    \STATE Take action $a_S$, update \textbf{\textit{$E_S$}}, observe $\textbf{\textit{O}}_S'$
    \STATE $f_d \gets \it{dtc}(E_S),~r_S  \gets \mathcal{R}(E_S, f_d)$
    \STATE $r_d\gets\mathcal{D}_{KL}(\pi_S ^{\theta_S}||\pi_T^{\theta_T})$
    \STATE Take action $a_t$, update \textbf{\textit{$E_S$}}, observe $\textbf{\textit{O}}_T'$
    \STATE $f'_d \gets \it{dtc}(E_T),~r_T \gets \mathcal{R}(E_T, f'_d)$
    \STATE $r'_d\gets\mathcal{D}_{KL}(\pi_T ^{\theta_T}||\pi_S^{\theta_S}$)
    \IF{$r_T \geq r_S$}
        \STATE $r_S^{rec} \gets r_S - \lambda_{S}\cdot (r_T - r_S) \cdot r_d$
        \STATE $r_T^{rec} \gets k \cdot r_T$
    \ELSIF{$r_T < r_S$}
        \STATE $r_S^{rec} \gets r_S + \lambda_{S}\cdot(r_S - r_T)$
        \STATE $r_T^{rec} \gets k \cdot r_T - \lambda_T \cdot (r_S - r_T) \cdot r'_d$
    \ENDIF
    \STATE $\delta_S \gets R_S^{rec}+\gamma\cdot V_S(\textbf{\textit{O}}_S';\omega_S)-V_S(\textbf{\textit{O}}_S;\omega_S)$
    \STATE $\delta_T \gets R_T^{rec}+\gamma\cdot V_T(\textbf{\textit{O}}_T';\omega_T)-V_T(\textbf{\textit{O}}_T;\omega_T)$
    \STATE $\omega_S \gets \omega_S - \alpha_S \cdot \delta_S \cdot \nabla_{\omega_S}V_S(\textbf{\textit{O}}_S;\omega_S) $
    \STATE $\omega_T \gets \omega_T - \alpha_T \cdot \delta_T \cdot \nabla_{\omega_T}V_T(\textbf{\textit{O}}_T;\omega_T) $
    \STATE $\theta_S \gets \theta_S - \beta_S \cdot \delta_S \cdot \nabla_{\theta_S}ln\pi_S(\textbf{\textit{O}}_S;\theta_S) $
    \STATE $\theta_T \gets \theta_T - \beta_T \cdot \delta_T \cdot \nabla_{\theta_T}ln\pi_T(\textbf{\textit{O}}_T;\theta_T) $
    \STATE $E_T \gets E_S$
\ENDWHILE
\ENDFOR
\end{algorithmic}
\end{algorithm}

%-------------------------------------------------------------------------------
% \section*{Acknowledgments}
% %-------------------------------------------------------------------------------

% The USENIX latex style is old and very tired, which is why
% there's no \textbackslash{}acks command for you to use when
% acknowledging. Sorry.

% %-------------------------------------------------------------------------------
% \section*{Availability}
% %-------------------------------------------------------------------------------

% USENIX program committees give extra points to submissions that are
% backed by artifacts that are publicly available. If you made your code
% or data available, it's worth mentioning this fact in a dedicated
% section.

%-------------------------------------------------------------------------------

%%%%%%%%%%%%%%%%%%%%%%%%%%%%%%%%%%%%%%%%%%%%%%%%%%%%%%%%%%%%%%%%%%%%%%%%%%%%%%%%
\end{document}